\documentclass[]{jfmnew}
\usepackage{graphicx}
\usepackage{array}
\usepackage{amsmath}
\usepackage{amsfonts}
\usepackage{pifont}
\usepackage[small,bf]{caption}
\usepackage{subfig}
\usepackage{tikz}
\usetikzlibrary{shapes,arrows}
\usetikzlibrary{calc}
\usepackage{pgf}
\usepackage{pgffor}
\usepackage{subfig}
\usepackage{caption}
\usepackage{hyperref}
\usepackage{rotating}
\usepackage[utf8]{inputenc}
\usepackage{amssymb}
\usepackage{color}
\usepackage{hyperref}
\usepackage[boxruled]{algorithm2e}
\usepackage{comment}
\usepackage{multirow}
\hypersetup{
	colorlinks=false,
	linktoc=all,
	linkcolor=blue,
	citecolor=blue,
}

\shorttitle{Sensor Selection for Performance Recovery}
\shortauthor{Huaijin Yao, Yiyang Sun, and Maziar Hemati}

\title{Feedback control of transitional shear flows: Sensor selection for performance recovery}

\author{Huaijin Yao,
  Yiyang Sun,
 \and Maziar S. Hemati \corresp{\email{mhemati@umn.edu}}}

\affiliation{Department of Aerospace Engineering and Mechanics, University of Minnesota,
Minneapolis, MN 55455, USA
}

\newcommand{\eqnref}[1]{(\ref{#1})}

\newcommand{\trans}{^\mathsf{T}}

\newcommand{\mR}{\mathbb{R}}

\definecolor{red2}{rgb}{0.8500, 0.1250, 0.0480}
\definecolor{green2}{rgb}{0.17, 0.66, 0.23}

\begin{document}

\maketitle

\begin{abstract}
The choice and placement of sensors and actuators is an essential factor determining the performance that can be realized using feedback control.
This determination is especially important, but difficult, in the context of controlling transitional flows. 
The highly non-normal nature of the linearized Navier-Stokes equations makes the flow sensitive to small perturbations, with potentially drastic performance consequences on closed-loop flow control performance.
Full-information controllers, such as the linear quadratic regulator~(LQR), have demonstrated some success in reducing transient energy growth and suppressing transition; however,  sensor-based output feedback controllers with comparable performance have been difficult to realize.
In this study, we propose two methods for sensor selection that enable sensor-based output feedback controllers to recover full-information control performance: one based on a sparse controller synthesis approach, and one based on a balanced truncation procedure for model reduction.
Both approaches are investigated within linear and nonlinear simulations of a sub-critical channel flow with blowing and suction actuation at the walls.
We find that sensor configurations identified by both approaches allow sensor-based static output feedback LQR controllers to recover full-information LQR control performance, both in reducing transient energy growth and suppressing transition.
Further, our results indicate that both the sensor selection methods and the resulting controllers exhibit robustness to Reynolds number variations.
\end{abstract}
\section{Introduction} \label{sec:introduction}
Preventing or delaying transition to turbulence via flow control is a topic of great technological interest. 
At a sufficiently high Reynolds number, a flow will transition from a low-skin-friction laminar regime to a high-skin-friction turbulent regime. 
For many wall-bounded shear flows, a sub-critical transition can arise due to non-modal instabilities~\citep{Schmid2001,Schmid2007}.
The high degree of non-normality of the linearized Navier-Stokes equations can cause flow perturbations to 
exhibit large peaks in kinetic energy, even when the flow is linearly stable.
This so-called \emph{transient energy growth~(TEG)} of flow perturbations serves as a driving mechanism for sub-critical transition.
TEG causes large deviations of the flow state from the laminar equilibrium, pushing it outside the basin of attraction and triggering secondary 
instabilities that ultimately transition the flow to turbulence~\citep{landahl1980,Brian1993,Reddy1993,Trefethen1993,jovanovic2005,Bamieh2001}.
An ability to reduce TEG---e.g.,~using flow control---could provide a means of suppressing transition to turbulence.

In studying TEG, it is important to note that not all flow perturbations will trigger transient growth, and so it is important to consider ``optimal'' or ``worst-case'' perturbations that 
give rise to the maximum TEG for a fixed perturbation amplitude~\citep{Butler1992}.
Considerations of worst-case performance are especially important when investigating and 
comparing the performance of different flow control strategies that aim to reduce TEG and suppress transition to turbulence.
The optimal disturbance for the uncontrolled flow will not necessarily be the same as the optimal disturbance for the controlled flow.
These optimal disturbances will vary further depending on the specific control design, 
and may even involve perturbations in the control system dynamics due to various sources of uncertainty~\citep{Hemati2018}.
Comparing worst-case performance ensures that comparisons of control performance remain fair.

Feedback control has been proposed as a way to reduce TEG and suppress transition in a number of flow configurations.
Within numerical simulations, the linear quadratic regulator~(LQR) has shown some success in reducing TEG and delaying transition in shear-flows using wall blowing and suction actuation~\citep{hogberg2003,Ilak2008,Martinelli2011,sun2019}.
LQR controllers are designed to minimize a balance between flow perturbation energy and control effort.
LQR control is an attractive choice because of its simplicity, but also because these controllers robustly reduce TEG in the face of parameter uncertainties (e.g.,~Reynolds number variations). 
Despite their appeal, LQR controllers require full-state information that tends to be unavailable outside of numerical simulations. 
Instead, sensor-based output feedback controllers are required to act based on partial information.
In principle, one can adopt an observer-based feedback approach, wherein measurements from a limited set of sensors are used to first estimate the full state of the flow, and to then apply a full-state feedback control law to these estimates.
Unfortunately, such approaches can exhibit degraded TEG performance if not properly designed~\citep{Hemati2018,Yao2018}.
Although sophisticated estimation strategies can be devised to overcome some of these performance issues~\citep{Bewley1998,hapffner2005}, 
these strategies have not been shown to recover full-information control performance. 

Static output feedback controllers have been proposed as a convenient design alternative to observer-based feedback control~\citep{Hemati2018,Yao2018}.
Static output feedback LQR~(SOF-LQR) control, in particular, can be designed using the same objective function as the full-information LQR controller, but with a constraint that the controller act as a direct feed-through from the measured sensor-outputs to commanded inputs~\citep{Toivonen1985,Syrmos1997,Cao1998a,Yao2019}.
This feed-through structure ensures that SOF-LQR controllers satisfy a necessary condition for TEG elimination~\citep{Hemati2018,Whidborne2007}.
In previous studies~\citep{Yao2018,Yao2019}, we have found that SOF-LQR using wall-based shear-stress sensing is able to reduce TEG, but as with LQG controllers, these control laws are unable to recover the same worst-case performance as the full-information LQR counterpart. 
This is not entirely surprising, as the achievable performance of a control law is intimately tied to the specific choice of actuators and sensors implemented for control.
Thus, the specific choice and arrangement of sensors is an important consideration in the design of sensor-based output feedback control strategies.

A number of studies have considered the problems of sensor selection for flow reconstruction~\citep{willcox2006,Manohar2018,Manohar2018OptimalSA,Clark2019,saito2020,yamada2020} and actuator selection for flow control~\citep{natarajan2016,chen2011,oehler2018,Bhattacharjee2020}.
Sensor selection for flow reconstruction has been of particular interest because it can benefit the design of flow estimation and diagnostic strategies.
However, in the context of TEG and transition control, even a ``perfect'' flow estimator can result in degraded control performance when the design is decoupled from the design of the control law~\cite{Hemati2018}.
Here, we are interested in selecting a set of sensors that enables sensor-based output feedback control to achieve comparable performance as a full-state feedback controller.
In particular, we aim to do this in the context of TEG reduction and transition control, without resorting to flow reconstruction and observer-based feedback designs.

In this paper, we propose two approaches for sensor selection that enable performance recovery of sensor-based output feedback controllers.
Both approaches are formulated as down-selection problems from a library of candidate sensors.
One approach leverages the fact that full-information LQR and SOF-LQR controllers can be made equivalent when the library of candidate sensors is sufficiently rich.
Thus, the controller gain for the associated SOF-LQR controller can be evaluated to determine the relative contribution of each candidate sensor to the control performance.
In doing so, sensors with little contribution to control actions can be identified and discarded to yield a sparse set of sensors without sacrificing flow control performance.
The second approach is based on ideas from linear model reduction using balanced truncation.
Balanced truncation yields a low-dimensional model that maintains the input-output dynamics of the system.
The low-order representation of the state-space for these reduced-order models also exposes redundant signals from sensors in the candidate library.
These redundant sensors can be identified and removed using a  pivoted QR decomposition to yield a sparse set of sensors for SOF-LQR controller synthesis.
We find that both of these approaches identify sensor configurations that allow SOF-LQR to recover full-state LQR TEG performance within linear simulations of a channel flow with wall-normal blowing and suction actuation. 
Each approach identifies a different configuration of velocity sensors distributed along the interior of the channel.
We further evaluate the control performance using direct numerical simulations~(DNS) of a nonlinear channel flow.
DNS are performed with various optimal disturbance amplitudes to examine the nonlinear effects and the control mechanism that delays or suppresses the laminar-to-turbulent transition.

The paper is organized as follows: In section~\ref{sec:model_synthesis}, we present the linearized channel flow model and an overview of relevant controller synthesis approaches.
In section~\ref{sec:sensor_selection}, we formulate two approaches for sensor selection that can be used for flow control performance recovery.
Section~\ref{sec:results} presents sensor configurations and linear performance analysis results for a sub-critical channel flow.
This is followed by nonlinear performance results from direct numerical simulations of a sub-critical  channel flow in section~\ref{sec:DNS}.
Finally, we draw conclusions in section~\ref{sec:conclusions}.

\section{Channel flow model and control synthesis}\label{sec:model_synthesis}
\subsection{Channel flow}
\label{sec:linearchannel}
\begin{figure}
\centering
\includegraphics[width=0.8\textwidth]{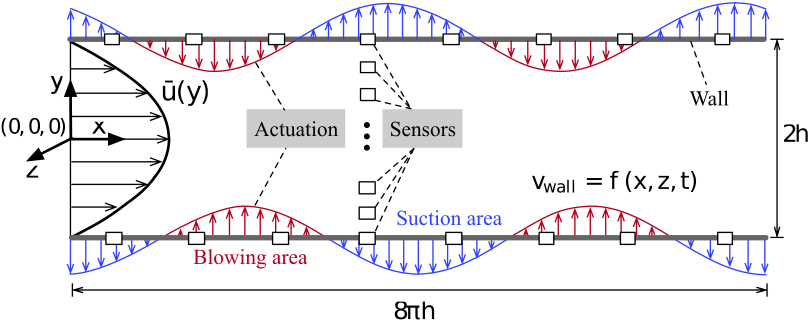}
\caption{Schematic of plane Poiseuille flow and implementation of sensing and actuation used for flow control (not to scale).} 
\label{fig:channel_flow}
\end{figure}

We consider plane Poiseuille flow at sub-critical Reynolds number of $Re= \boldsymbol{\bar u_c} h/\nu=3000$ defined by the centerline velocity of the base flow $\boldsymbol{\bar u_c}$, half height of the channel $h$, and kinematic viscosity $\nu$. As shown in figure \ref{fig:channel_flow}, the velocity profile of the base flow is $[\boldsymbol{\bar u}, \boldsymbol{\bar v}, \boldsymbol{\bar w}]=[1-(y/h)^2,0,0]$ with coordinate origin at the center line between walls, where $\boldsymbol{\bar u},\boldsymbol{\bar v}, \boldsymbol{\bar w}$ represent velocity components in streamwise $x$, wall-normal $y$, and spanwise $z$ directions, respectively. The parabolic profile of the velocity is the laminar equilibrium solution of the flow. In this study, length-scale variables are non-dimensionalized by the channel half-height $h$, and velocities are non-dimensionalized by the centerline velocity of base flow $\boldsymbol{\bar u_c}$. Time is denoted by $t$.   

The flow at this condition is linearly stable, since there are no unstable modes from the linear stability analysis. However, a large transient energy growth of small perturbations is observed at this flow condition, and a laminar-to-turbulent transition emerges as a result of certain flow perturbations. Hence, our objective is to design feedback control strategies to suppress the transient energy growth and further prevent the emergence of laminar-to-turbulent transition. 

The flow control configuration is illustrated in figure \ref{fig:channel_flow}. We introduce actuation in the form of blowing and suction in the wall-normal direction on the upper and lower channel walls. The velocity profile of the actuation is spatially periodic in the streamwise and spanwise directions, in accordance with the channel flow model discussed below. In this study, we consider velocity sensors along the channel interior, but the methods we introduce are valid equally with wall-based sensing as well.

\subsection{Linearized Navier--Stokes equations} 
\label{sec:fluid_model}
We decompose the flow state $\boldsymbol{q}$ into base state $\bar {\boldsymbol{q}}$ and small perturbation $\boldsymbol{q}'$ ($\boldsymbol{q}=\bar {\boldsymbol{q}}+\boldsymbol{q}'$), where $\boldsymbol{q}=[\boldsymbol{u}, \boldsymbol{v},\boldsymbol{w},\boldsymbol{p}]\trans$ ($\boldsymbol{p}$ is pressure), and $(\cdot){\trans}$ represents transpose. The kinetic energy density of a perturbation is defined as, 
\begin{equation}
E=\frac{1}{2V}\int_{V} (\boldsymbol{u}'^2+\boldsymbol{v}'^2+\boldsymbol{w}'^2) dV,
\end{equation}
where $V$ is the volume of the computational domain.

By substituting the expression of $\boldsymbol{q}=\bar{\boldsymbol{q}}+\boldsymbol{q}'$ into the Navier--Stokes equations, and assuming that the perturbation is much smaller than the base state in magnitude ($|\boldsymbol{q}'|\ll |\bar{\boldsymbol{q}}|$), we linearize the equations by retaining linear terms and neglecting higher-order nonlinear terms as follows,
\begin{equation}
\begin{split}
\frac{\partial \boldsymbol{u}'}{\partial x} + \frac{\partial \boldsymbol{v}'}{\partial y} + \frac{\partial \boldsymbol{w}'}{\partial z}~~~~&=0\\
\frac{\partial \boldsymbol{u}'}{\partial t} + \boldsymbol{\bar u} \frac{\partial \boldsymbol{u}'}{\partial x} + \boldsymbol{v}'\frac{\partial\boldsymbol{\bar{u}}}{\partial y} &=\frac{\partial \boldsymbol{p}'}{\partial x} +\frac{1}{Re} \nabla ^2 \boldsymbol{u}'\\
\frac{\partial \boldsymbol{v}'}{\partial t} + \boldsymbol{\bar{u}} \frac{\partial \boldsymbol{v}'}{\partial x} ~~~~~~~~~~~ &=\frac{\partial \boldsymbol{p}'}{\partial y} +\frac{1}{Re} \nabla ^2 \boldsymbol{v}'\\
\frac{\partial \boldsymbol{w}'}{\partial t} + \boldsymbol{\bar u}  \frac{\partial \boldsymbol{w}'}{\partial x} ~~~~~~~~~~ &=\frac{\partial \boldsymbol{p}'}{\partial z} +\frac{1}{Re} \nabla ^2 \boldsymbol{w}'.\\
\end{split}
\label{LN}
\end{equation}

The linearzied Navier--Stokes equations are further manipulated to be expressed in terms of wall-normal velocity $\boldsymbol v'$ and wall-normal vorticity $\boldsymbol \eta'$ as described in~\citep{Schmid2001} as, 
\begin{equation}
   \begin{split}
\frac{\partial{(\nabla^2 \boldsymbol{v'})}}{\partial t}+ \boldsymbol{\bar u}\frac{\partial{(\nabla^2\boldsymbol{v'})}}{\partial x}-\frac{\partial^2 \bar{\boldsymbol{u}}}{\partial y^2}\frac{\partial \boldsymbol{v'}}{\partial x}-\frac{1}{Re}\nabla^2(\nabla^2 \boldsymbol{v'}) &= 0 \\
\frac{\partial \boldsymbol{\eta'}}{\partial t}+\frac{\partial \bar{\boldsymbol{u}}}{\partial y}\frac{\partial \boldsymbol{v'}}{\partial z}+\bar{\boldsymbol{u}}\frac{\partial \boldsymbol{\eta'}}{\partial x}-\frac{1}{Re}\nabla^2\boldsymbol{\eta'} ~~~~~~~~~~~~~~~~~~~ &=0.
\end{split} 
\end{equation}
Next, the real-valued three-dimensional perturbation of wall-normal velocity and wall-normal vorticity are expressed using Fourier expansions in the homogeneous $x$- and $z$-directions,   
\begin{equation} 
\begin{split}
\boldsymbol{v'}(x,y,z,t) = \hat {\boldsymbol{v}}(y,t)e^{i(\alpha x +\beta z)} + \text{complex conjugate},\\
\boldsymbol{\eta'}(x,y,z,t) = \hat {\boldsymbol{\eta}}(y,t)e^{i(\alpha x +\beta z)} + \text{complex conjugate},
\label{modal}
\end{split}
\end{equation}
where $\hat {\boldsymbol{v}}(y,t)$ and $\hat {\boldsymbol{\eta}}(y,t)$ are amplitude functions of the perturbation associated with streamwise wavenumber $\alpha$ and spanwise wavenumber $\beta$. By plugging~(\ref{modal}) into the linearized Navier--Stokes equations (\ref{LN}), we obtain the Orr-Sommerfeld and Squire equations as follows,
\begin{equation}
\left[\begin{array}{c}
    \dot{\hat{\boldsymbol{v}}}  \\
     \dot{\hat{\boldsymbol{\eta}}}
\end{array} \right]
     =
     \left[\begin{array}{cc}
   - i \alpha \boldsymbol{\bar u}+\frac{i\alpha \boldsymbol{{\bar u}}_{yy}}{(\mathcal{D}^2-k^2)}+ \frac{1}{Re} (\mathcal{D}^2-k^2) & 0  \\
     -i\beta \boldsymbol{{\bar u}}_y & - i\alpha \boldsymbol{\bar u}+\frac{1}{Re}(\mathcal{D}^2-k^2)
     \end{array}\right]
     \left[\begin{array}{c}
    {\hat{\boldsymbol{v}}}  \\
    {\hat{\boldsymbol{\eta}}}
\end{array} \right],
\label{eqn:v_v}
\end{equation} 
where $\mathcal{D}$ represents differentiation with respect to the wall-normal direction $y$, $\boldsymbol{{\bar u}}_y$ and $\boldsymbol{\bar u}_{yy}$ denote the first and second derivatives of $\boldsymbol{\bar u}$ with respect to $y$, and $k^2 := \alpha^2+\beta^2$. 
The governing equations~(\ref{eqn:v_v}) are in a state-space form 
\begin{equation}
\frac{\partial X_u(y)}{\partial t} = {A_u}(\boldsymbol{\bar q}; \alpha, \beta)  {X_u(y)},
\label{OL}
\end{equation} 
where ${X_u} = [\hat {\boldsymbol{v}}, \hat {\boldsymbol{\eta }}]\trans$ is the state, and subscript $(\cdot)_u$ indicates uncontrolled system. 

In the $y$-direction, flow variables are represented by Chebyshev polynomials with $N=101$ discrete collocation points, and a no-slip boundary condition is prescribed at the upper and lower walls for the uncontrolled baseline flow: i.e., $\hat{\boldsymbol v}(\pm h)={\hat{\boldsymbol v}}_y(\pm h)=\hat{\boldsymbol{\eta}}(\pm h)=0$.

In the control design, we introduce actuation in the form of wall-normal blowing and suction at the upper and lower channel walls (see figure \ref{fig:channel_flow}). This actuation modifies the uncontrolled dynamic system~(\ref{OL}) to form a controlled dynamical system
\begin{equation}
\frac{\partial X}{\partial t}  =  AX + BU,
\label{eqn:ss_CL}
\end{equation}
where $A$ is the system matrix, $X$ is the state, $U$ is the input, and ${B}$ is the input matrix that maps the influence of control inputs to the state evolution. 
Here the control input is selected to be ${U}=\frac{\partial}{\partial t}[\hat {\boldsymbol v}|_{+h}, \hat {\boldsymbol v}|_{-h}]^T$, representing the rate of change of wall-normal velocity on the upper and lower walls. 
Since the control input $U$ represents the change of Fourier coefficients of wall-normal velocity $\hat {\boldsymbol v}$, in figure~\ref{fig:channel_flow} it is shown in the form of a sinusoidal wave. The no-flow-through boundary conditions ($\hat {\boldsymbol v}|_{\pm h}=0$) are excluded from $X_u$ in~\eqref{OL}. In the controlled case, $\hat {\boldsymbol v}|_{\pm h}$ are nonzero, so we append these two state variables to the flow state $X_u$ to form a new state $X=[X_u, \hat {\boldsymbol v}|_{+h}, \hat {\boldsymbol v}|_{-h}]\trans$. 
Analogously, the dynamics matrix $A_u$ is modified and denoted by $A$ to account for the new state. 

In all that follows, we transform all quantities to their equivalent real-valued representations so that $A \in \mathbb{R}^{n \times n}$, $X \in \mathbb{R}^{n}$, ${B}\in \mathbb{R}^{n \times m}$, and $U \in \mathbb{R}^{m}$.
Further, all measured sensor outputs $Y\in\mathbb{R}^p$ to be considered in this study will be represented by the output equation
\begin{equation}
Y(t) = CX(t).  \label{eqn:output}
\end{equation}
Further details about the model formulation can be found in~\citep{McKernanIJMIC2006}.
\subsection{Transient energy growth and control synthesis}
In this study, we aim to use feedback control to reduce the transient energy growth~(TEG). We consider TEG due to some initial flow perturbation $X(t_0)=X_0$. 

The associated system response is given by $X(t)=\mathrm{e}^{A(t-t_0)}X_0$, and the associated perturbation kinetic energy is given as,
\begin{equation}
E(t)=X\trans(t)QX(t),  \label{eqn:Q_define}
\end{equation}
where $Q=Q\trans > 0$. 
Further, the maximum TEG is defined as,
\begin{equation}
G=\max _{t \geq t_0}\max _{E(t_0)  \neq 0} \frac{E(t)}{E(t_0)}, % \; Q=Q^T > 0 
\end{equation}
which results from a so-called \emph{worst-case} or \emph{optimal perturbation}~\citep{butlerPOF1992}.
The system can exhibit TEG whenever $G>1$.
In this study, we always evaluate the maximum TEG for a given system. The worst-case perturbations are calculated using the algorithm proposed in~\citep{Whidborne2011}.
We emphasize that, in general, perturbation for the uncontrolled flow will be different from the one for the controlled flow, and so these perturbations must be determined independently.

To use feedback control to achieve TEG reduction, the control input vector $U$ at a given instant is determined from available system information. For a full-state feedback control, the state $X$ is assumed to be known and available for feedback, i.e., full-information control:
\begin{equation}
    U(t)=KX(t), \label{eqn:state_input}
\end{equation}
where the design variable $K\in\mR^{m\times n}$ is called the state feedback gain matrix.

LQR synthesis is based on solving,
\begin{equation}
\min_{U(t)} J = \int _0 ^\infty X\trans(t)QX(t)+U\trans(t)RU(t)dt \label{eqn:objective_function}
\end{equation}
subject to the linear dynamic constraint given in~(\ref{eqn:ss_CL}), where $R>0$. 
The resulting full-state feedback LQR controller gain matrix $K$ is determined from the solution of an algebraic Riccati equation~\citep{Brogan1991}. 
Although LQR controllers will not necessarily minimize TEG, they have been shown to reduce TEG in shear flows and to exhibit robustness to parametric uncertainties~\citep{Ilak2008,Martinelli2011,sun2019}.

Outside of numerical simulations, full-state feedback controllers are typically not practically viable for flow control; such controllers require knowledge of the full state of the flow, which is usually not directly available for feedback in practice.
In order to achieve feedback control with the measured information from sensors, in this study, we propose to use a static output feedback~(SOF) control structure. For SOF control, the control input is determined directly from the measured output $Y$ in an analogous manner to full-state feedback, with
\begin{equation}
    U(t)=FY(t), \label{eqn:output_input}
\end{equation}
where the design variable $F\in\mR^{m\times p}$ is called the SOF feedback gain matrix.
The SOF form of LQR control was proposed for TEG control in~\citep{Yao2018,Yao2019}.%\ys{check the year of reference}.
The SOF-LQR solves the same minimization problem as shown in~\eqnref{eqn:objective_function}, but with an SOF constraint on the feedback law. This is equivalently written as, 
\begin{equation}
J=\int _0 ^ \infty X\trans(t)[Q+(FC)\trans R(FC)]X(t)dt.
\end{equation}
The SOF-LQR can be solved using iterative Anderson-Moore methods~\citep{Rautert1997,Syrmos1997}. In this study, we use the Anderson-Moore algorithm with Armijo-type adaptation proposed in~\citep{Yao2018}. The details are given in Appendix A.

\section{Sensor selection for output feedback performance recovery} \label{sec:sensor_selection}

In this section, we propose two methods for selecting 
a sparse set of sensors to enable sensor-based SOF 
controllers to recover full-information control performance.
We restrict the discussion to sensor selection and linear quadratic performance, but the approach is applicable more generally and can be used for actuator selection and other performance measures just as well.

The basic idea begins by recognizing that SOF-LQR control
can be made equivalent to full-information LQR control 
if an available set of $p\ge n$ sensors is \emph{sufficiently rich}.
Here, the term \emph{sufficiently rich} amounts to requiring the output matrix
to have full column rank (i.e.,~$C=C_n\in\mathbb{R}^{p\times n}$ with $\mathrm{rank}(C_n)=n$).
Since the closed-loop dynamics for SOF-LQR control are given by
\begin{equation}
  \dot{X}(t)=(A+BF_nC_n)X(t), \label{eqn:sys_fullC}  
\end{equation}
it follows that the SOF-LQR gain $F_n$ can be designed from the full-state feedback LQR gain $K$ to recover full-state feedback LQR performance exactly: simply find a gain $F_n$ that satisfies $F_nC_n=K$.
From this insight, it follows that if we form a sufficiently rich library of candidate sensors, then the sensor selection problem can be recast as a problem of sensor down-selection.
That is, given a sufficiently rich library of sensors, we aim to determine 
which sensors can be removed from the library such that the closed-loop performance of the controlled system~\eqref{eqn:sys_fullC} will be minimally impacted.
Doing so will result in a sparse set of $r < n$ sensors with associated output matrix $C_r \in \mathbb{R}^{r \times n}$.

Here, we propose two methods for performance recovery via sensor down-selection from a sufficient rich library of candidate sensors: one based on controller gain evaluation (see section~\ref{sec:approach_CS}), and another based on a balanced truncation procedure (see section~\ref{sec:approach_BT}).
Both methods will make it possible for SOF-LQR controllers to recover full-state LQR performance, as will be demonstrated in sections~\ref{sec:results} and \ref{sec:DNS}.

\subsection {Sensor Selection by column-norm evaluation (CE)} \label{sec:approach_CS}

The first approach we propose stems from the fact that a controller gain matrix contains important information regarding the relative contribution of individual sensors and actuators to the controlled closed-loop dynamics in~\eqref{eqn:sys_fullC}.
If we design a controller gain~$F_n$ based on a sufficiently rich library of candidate sensors, then
each column in~$F_n$ corresponds to a specific sensor in the candidate library, represented as a row of~$C_n$.
Thus, by evaluating the relative norm of each column in~$F_n$, we can determine the relative contribution of a particular sensor to the control action and the closed-loop dynamics in~\eqref{eqn:sys_fullC}.
Here, we assess the relative importance of sensors by evaluating the relative $L_2$-norm of each column in the gain matrix~$F_n$.
Let $F_{n}(:,i)$ indicate the~$i^{th}$ column of~$F_n$, then the $L_2$-norm is calculated as
\begin{equation}
    \|F_{n}(:,i)\|_2 = \sqrt{\sum ^n _{j=1} F_{n}(j,i)^2}, \label{eqn:normF}
\end{equation} 
where $F_{n}(j,i)$ is the $j^{th}$ element of the $i^{th}$ column of $F_n$.
Consider, for example, that a column in $F_n$ of all zeros and the associated row in $C_n$ could be removed completely without altering the closed-loop system response in~\eqref{eqn:sys_fullC}. 
In general, columns of $F_n$ with large norm contribute more to the closed-loop response than the columns of $F_n$ with small norm.
Thus, the rows in $C_n$  associated with the dominant columns in $F_n$ indicate sensors that are ``more important'' for the controlled system dynamics.
Therefore, we identify the subset of $r$ columns with the largest relative norm, denoted by their indices ${j_1,\cdots, j_r}$.
Then, a new output matrix $C_r$ containing a reduced set of $r$ sensors can be constructed
from the rows ${j_1,\cdots, j_r}$ of $C_n$.
In principle, a gain matrix $F_r$ for this reduced set of sensors can be determined directly from the columns ${j_1,\cdots, j_r}$ of $F_n$, but it is actually beneficial to compute an optimal gain $F$ by re-designing the SOF gain for the system $(A,B,C_r)$.
In the present work, we consider linear quadratic control objectives, so the re-design is performed based on SOF-LQR synthesis via an iterative Anderson-Moore algorithm (see Appendix~\ref{appA}). 

The method proposed here for sensor selection by column-norm evaluation~(CE) is summarized in Algorithm~\ref{Algorithm2}.
Note that it can be useful to scale the outputs of the candidate sensor library so that each row of $C_n$ has unit norm.
This results from the fact that the scaling of the sensor output is inversely associated with the scale of the associated column norm in $F_n$.
Performing this scaling to unit norm output tends to be important when a heterogeneous set of sensors is used to construct the candidate library (e.g.,~velocity, pressure, and shear-stress);
otherwise, the importance of a particular type of sensor can be artificially inflated or deflated by the nature of observable quantity itself.

Lastly, note that a similar procedure to Algorithm~\ref{Algorithm2} can be formulated for actuator selection by considering the relative norm of rows in $F_n$, each corresponding to an actuator represented as a column of the input matrix~$B$ in~\eqref{eqn:sys_fullC}.
These ideas are closely related to sparse controller synthesis techniques based on convex optimization, which can be used to explicitly promote sparsity and design controller gains with many columns or rows of all zeros~\citep{Lin2013,Polyak2014}.

\begin{center}
\begin{minipage}[t]{0.85\textwidth}
\begin{algorithm}[H] \label{Algorithm2}

\textbf{step 0:} Form a sufficiently rich library of candidate sensors~$C_n$, and design an SOF controller $F_n$ that achieves the same performance as the desired full-information controller by solving $F_nC_n=K$.

\textbf{step 1:} Evaluate the $L_2$-norm of each column in $F_n$, then save the indices $\{j_1,\cdots, j_r\}$ of the $r$ columns with the largest relative $L_2$-norm.

\textbf{step 2:} Construct a reduced measurement matrix $C_r$ whose $r$ rows consist of rows ${j_1,\cdots,j_r}$ of $C_n$.

\textbf{step 3:} Re-design an SOF controller based on $(A,B,C_r)$.

\caption{\textbf{\textit{Sensor selection by column-norm evaluation~(CE)}}}

\end{algorithm}
\end{minipage}
\end{center}

\subsection {Sensor selection by balanced truncation} \label{sec:approach_BT}

Another sensor selection procedure can be devised using the notion of balanced truncation~\citep{Moore1981,Antoulas2005,Laub1987}. 
Balanced truncation is a model reduction method for linear systems that
works by truncating states with a lesser contribution to a system's input-output dynamics.
To do so, the state-space is first transformed into balanced coordinates $\bar{X}_b=TX$, so that the controllability Gramian $W_c$ and the observability Gramian $W_o$ are equal and diagonal.
That is, in balanced coordinates we have $\bar{W}_c=\bar{W}_o=diag(\sigma_1,\cdots,\sigma_n)$,
where $\sigma_1 \geq \sigma_2\geq \cdots \ge\sigma_n$ are the system's Hankel Singular Values~(HSVs).
Larger HSVs indicate directions of state-space with greater contribution to the input-output dynamics, whereas smaller HSVs indicate directions of state-space with lesser contribution to the input-output dynamics.
As such, truncation in balanced coordinates of the $n-r$ states with the smallest HSVs will allow a reduction to an $r$-dimensional state-space, while preserving information that is most important for capturing the input-output dynamics.
This idea can be extended for sensor selection from a sufficiently rich library of candidate sensors $C_n$, since a reduction to an $r$-dimensional state-space will necessarily lead to an output matrix with redundant rows.
Upon eliminating these redundant rows in $C_n$---e.g.,~by means of a pivoted QR decomposition---we will be left with $r$ linearly independent rows (sensors) that are relevant for feedback control.

\begin{figure}
\begin{centering}
\vspace{2pt}
\subfloat[Original system with full rank measurements]{\label{fig:BT_approach1}
\includegraphics[height=1.5cm]{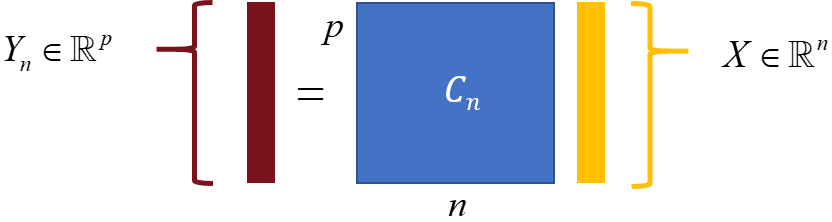}
} 
\hspace{1pt}
\subfloat[Input-output behavior preserved by balanced truncation with partial state]{\label{fig:BT_approach2}
\includegraphics[height=1.5cm]{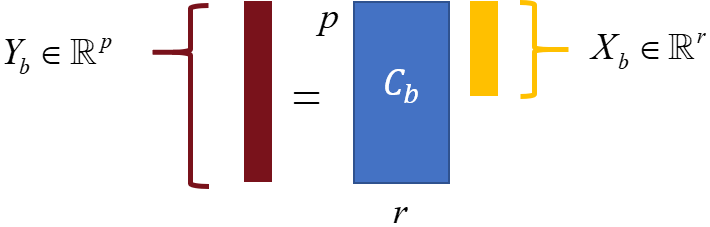}
} \\
\vspace{1pt}
\hspace{1pt}
\subfloat[Keep an independent set of measurements]{\label{fig:BT_approach3}
\includegraphics[height=1.5cm]{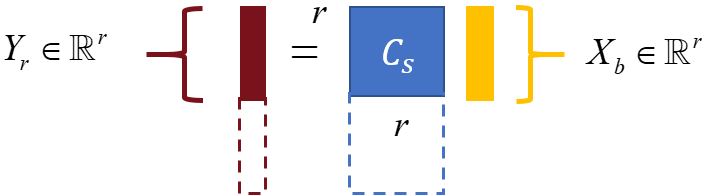}
}
\hspace{1pt}
\subfloat[Full states with reduced measurements]{\label{fig:BT_approach4}
\includegraphics[height=1.5cm]{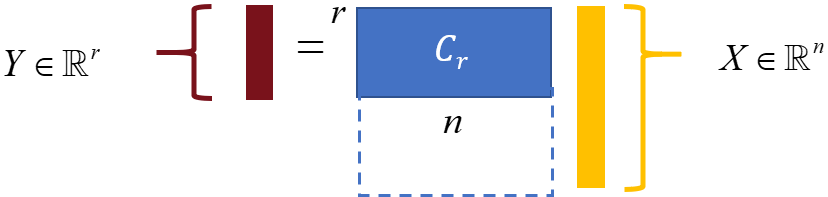}
}
\caption{Illustration of measurement selection by balanced truncation.} \label{fig:BT_approach}
\end{centering}
\end{figure}

The balancing transformation can be determined by first computing the system Gramians from the associated Lyapunov equations,
\begin{equation}
    \begin{aligned}
    AW_c+W_cA^T+BB^T &=0\\
    W_oA+A^TW_o+C_n^TC_n &=0. \label{eqn:gramians}
    \end{aligned}
\end{equation}
Then, using the lower triangular Cholesky factorizations $L_o$ and $L_c$ of $W_o$ and $W_c$, respectively, and the singular value decomposition
of their product $L_o^TL_c = \Phi\Sigma\Psi^T$, the balancing transformation can be computed as
\begin{equation}
\begin{aligned}
    T ~~~&=L_c\Psi\Sigma^{\frac{1}{2}} \\
    T^{-1}&=\Sigma^{\frac{1}{2}}\Phi^TL_o^T. \label{eqn:balanced_coor}
\end{aligned}
\end{equation} 
By retaining the $r$ states with the largest HSVs and truncating the rest, the reduced-order system resulting from balanced truncation can be expressed as
\begin{equation}
\begin{aligned}
\dot{X}_b(t) &= A_b X_b(t) +B_b U(t) \\
Y_b(t) &= C_b X_b(t), 
\end{aligned} \label{eqn:BC_sys}
\end{equation} 
where $A_b \in \mathbb{R}^{r \times r}$, $B_b \in \mathbb{R}^{r \times m}$, $C_b \in \mathbb{R}^{p \times r}$, the reduced-order state $X_b \in \mathbb{R}^{r}$, input vector $U \in \mathbb{R}^{m}$, and output vector $Y_b \in \mathbb{R}^{p}$.
Since $r<n$, the output matrix $C_b$ will necessarily have redundant rows.
Thus, only $r$ linearly independent rows (sensors) are needed to achieve the same feedback control performance as the reduced-order system with the full sensor library.
To see this, consider that the static output feedback control determines the control action directly from the measured output as $U=F_nY_n$.
This can be approximately achieved based on the reduced-order model attained via balanced truncation as $U=F_nY_n \approx F_nY_b$ (see figure~\ref{fig:BT_approach}).
Now, since $Y_b=C_bX_b$ and $C_b$ has redundant rows,
it is possible to exactly reproduce this control signal from a reduced set of $r$ outputs $Y_r=C_sX_b$ with $C_s\in\mathbb{R}^{r\times r}$ as $U=F_nY_b=F_rY_r$.
This final output signal can be represented in the original basis for the full $n$-dimensional state-space, which determines the specific sensors that are needed to recover full-information feedback control performance using output feedback control.

In order to down-select to a set of $r$ linearly independent rows from the output matrix $C_b$, we make use of the column-pivoted QR decomposition as 
\begin{equation}
    C_b\trans \mathbf{E} = \mathbf{Q}\mathbf{R},
    \label{eq:qrdecomp}
\end{equation}
where $\mathbf{Q} \in \mathbb{R}^{r \times r}$ is an orthogonal matrix and $\mathbf{R} \in \mathbb{R}^{r \times p}$ is an upper triangular matrix with diagonal elements $r_{ii}$, with $|r_{11}|\ge|r_{22}|\ge\dots\ge|r_{nn}|$. 
The independent rows $j_1,\cdots,j_r$ of $C_b$ can be identified by entries in the column permutation matrix $\mathbf{E} \in \mathbb{R}^{p \times p}$.
Thus, only these rows of $C_b$ are retained in order to down-select the number of sensors and obtain the associated output matrix $C_s$ shown graphically in figure~\ref{fig:BT_approach3}.
Similarly, these same indices can be used to construct the output matrix $C_r$ in the original $n$-dimensional basis for the state-space: simply construct $C_r$ from rows $j_1,\cdots,j_r$ of $C_n$.
All that remains at this point is to re-design an SOF controller based on the system $(A,B,C_r)$ that uses this reduced set of sensors.
A summary of this balanced truncation~(BT) sensor selection approach is summarized as Algorithm \ref{Algorithm4} below.
Note that, as with the CE sensor selection approach, this BT approach can be formulated analogously for actuator selection as well.
Similar ideas for using balanced model reduction methods for sensor and actuator selection have been investigated in~\citep{Manohar2018OptimalSA}.

The choice of $r$ in the carrying out the above steps to is important for ensuring that the final sensor configuration is able to achieve the desired control performance.
One way to make this determination is to evaluate the 
difference between the input-output dynamics of the $r$-dimensional reduced-order balanced truncation model relative to the $n$-dimensional full-order model.
We can evaluate this error by first defining an error system $\Sigma_e$ that maps any input $U$ to the associated error at the output $e=Y_n-Y_b$:
\begin{equation}
\begin{aligned}
\left[ \begin{array}{c}
    \dot{X}(t) \\
     \dot{X}_b(t) 
\end{array}\right] &=  \left[ \begin{array}{cc}
    A  &  ~~~0  \\
    0 & ~~~A_b
    \end{array}\right] \left[ \begin{array}{c}
    X(t) \\
     X_b(t) 
\end{array}\right]+
    \left[ \begin{array}{c}
    B  \\
    B_b
    \end{array}\right] U(t)
    \\
e(t) ~~~~~&=  \left[\begin{array}{cc}
    C_n  & -C_b  
\end{array}\right] \left[ \begin{array}{c}
    X(t) \\
     X_b(t) 
\end{array}\right].
\end{aligned} \label{eqn:err_sys}
\end{equation} 
Then, an appropriate $r$ can be determined by evaluating an associated system norm of $\Sigma_e$ and ensuring it falls below some threshold~$err$.
A small system norm When the error system holds small system norms, the ROM captures the original system's input-output dynamics well.
In this study, we can consider both the $\mathcal{H}_2$- and $\mathcal{H}_{\infty}$-norms of $\Sigma_e$.
The $\mathcal{H}_2$-norm of $\Sigma_e$ is denoted as $\Vert \Sigma_e \Vert _{\mathcal{H}_2}$ and corresponds to the root-mean-square of the impulse response of $\Sigma_e$.
The $\mathcal{H}_\infty$-norm of $\Sigma_e$ is denoted by $\Vert \Sigma_e \Vert _{\mathcal{H}_{\infty}}$ and corresponds to the peak value of the largest singular value of $\Sigma_e$.

\begin{center}
\begin{minipage}[t]{0.85\textwidth}
\begin{algorithm}[H] \label{Algorithm4}

\textbf{step 0:} Form a sufficiently rich library of candidate sensors~$C_n$.

\textbf{step 1:} Transform the system~$(A,B,C_n)$ to balanced coordinates using the balancing transformation in~\eqref{eqn:balanced_coor}.

\textbf{step 2:} Perform a balanced truncation to the $r$-dimensional system representation in~\eqref{eqn:sys_fullC}, with associated output matrix $C_b$.

\textbf{step 3:} Perform a column-pivoted QR factorization of $C_b\trans$ as in~\eqref{eq:qrdecomp} to identify a set of $r$ linearly independent rows $\{j_1,\cdots, j_r\}$ from $C_b$.

\textbf{step 4:} Construct a new output matrix $C_r$ from rows $\{j_1,\cdots, j_r\}$ of the original candidate library of sensors $C_n$.
This $C_r$ represents the sparse sensor configuration.

\textbf{step 5:} Design an SOF controller based on the system~$(A,B,C_r)$.

\caption{\textbf{\textit{Sensor selection by balanced truncation~(BT)}}}
\end{algorithm}
\end{minipage}
\end{center}  

\section{Results: sensor selection and linear performance analysis} \label{sec:results}

In the previous section, we proposed two approaches for sensor selection---one based on column-norm evaluation~(CE) and one based on balanced truncation~(BT)---that can enable a sensor-based SOF controller to recover full-information control performance.
Here, we apply each method for sensor-based output feedback control of the linearized channel flow with $Re=3000$ described in section~\ref{sec:linearchannel}.
The SOF-LQR controllers here will be designed to recover the full-information LQR control TEG performance subject to worst-case streamwise $(\alpha,\beta)=(1,0)$, oblique $(\alpha,\beta)=(1,1)$, and spanwise $(\alpha,\beta)=(0,2)$ disturbances.
Sensors will be determined from a sufficiently rich library of velocity sensors, distributed at Chebyshev collocation points along the wall-normal direction throughout the interior of the channel, but excluding points nearest to the walls (see figure~\ref{fig:channel_flow}). 
Since the flow is spatially invariant, the sensor library is constructed based on Fourier coefficients.
For streamwise disturbances, only wall-normal velocity information is
relevant and so a sufficiently rich sensor library in this case is built-up
using $\hat {\boldsymbol v}$ sensors only.
For oblique and spanwise disturbances, we use $\hat {\boldsymbol u}$ and $\hat {\boldsymbol v}$ to build up our sensor library, so that  $\mathrm{rank}(C_n)=n$.
We emphasize that these choices in the construction of the sensor library
are not unique, and are selected here mainly for simplicity.

\begin{figure}
%\vspace{2pt}
\subfloat[$(\alpha,\beta)=(1,0)$]{\label{fig:normF_desend_a1b0}
\includegraphics[width=0.31\textwidth]{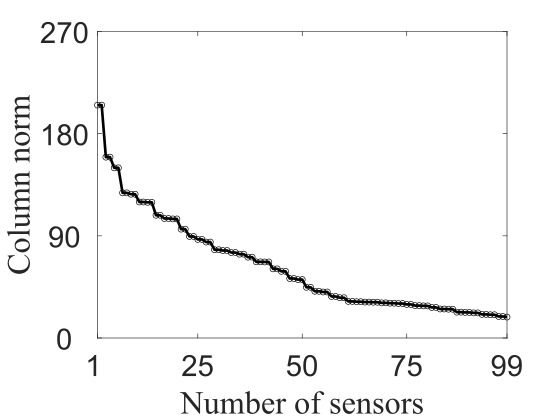}
}
\hspace{0.05pt}
\subfloat[$(\alpha,\beta)=(1,1)$]{\label{fig:normF_desend_a1b1}
\includegraphics[width=0.31\textwidth]{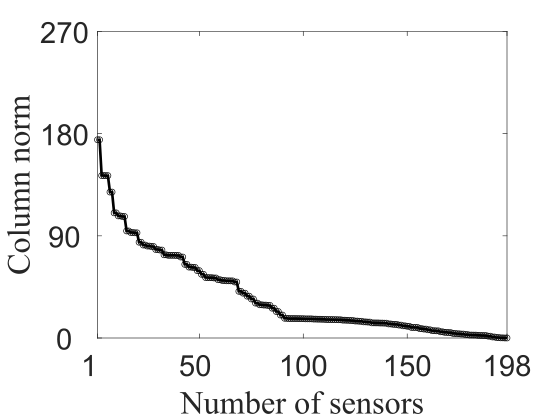}
}
\hspace{0.05pt}
\subfloat[$(\alpha,\beta)=(0,2)$ ]{\label{fig:normF_desend_a0b2}
\includegraphics[width=0.31\textwidth]{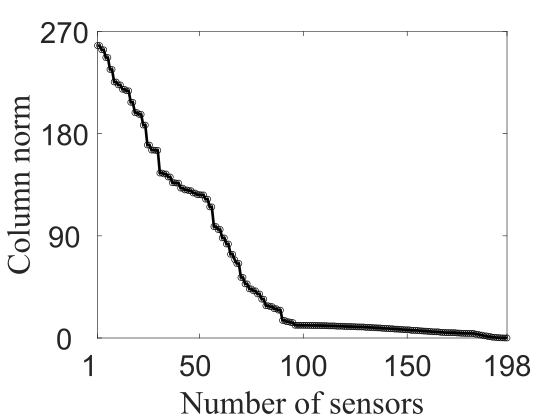}
} 
\caption{In the gain column norm evaluation approach, evaluate the column norm of the gain matrix $F_n$. The L2-norms of each column are sorted in descend order. } \label{fig:normF_descend}
\end{figure}

In carrying out the CE approach (Algorithm~\ref{Algorithm2}), we begin by evaluating the $L_2$-norm for each column of the controller gain $F_n$, then rank these from largest to smallest (see figure~\ref{fig:normF_descend}).
For each of the three wavenumber combinations reported in figure~\ref{fig:normF_descend}, there are a few columns with a relatively large $L_2$-norm compared to the others.
Recall that a larger norm indicates that a sensor has a more dominant contribution to the control action, and so this indicates that the corresponding sensors should be retained, while sensors associated with smaller norms can be truncated.
We do this in sequence, truncating all but the largest norm sensors, then re-designing the SOF-LQR controller.
We compare the worst-case TEG for the new SOF-LQR controller with that of the full-information LQR.
If the worst-case performance is not satisfactorily matched between the two, then we proceed to introduce the next set of dominant sensors according to the rankings in figure~\ref{fig:normF_descend}, and repeat this process until the resulting SOF-LQR controller recovers full-information LQR controller performance.
Each graph of column norm versus number of sensors $r_{CE}$ in figure~\ref{fig:normF_descend} exhibits at least one ``elbow point'', and so we expect that we will be able to recover full-information LQR performance using a sparse sensor configuration with SOF-LQR control.
Indeed, this will be the case.
These results will be presented and discussed in more detail after we discuss the preliminary aspects of the BT procedure for sensor selection.

\begin{figure}
%\vspace{2pt}
\subfloat[$(\alpha,\beta)=(1,0)$]{\label{fig:a1b0_error_sysnorm}
\includegraphics[width=0.31\textwidth]{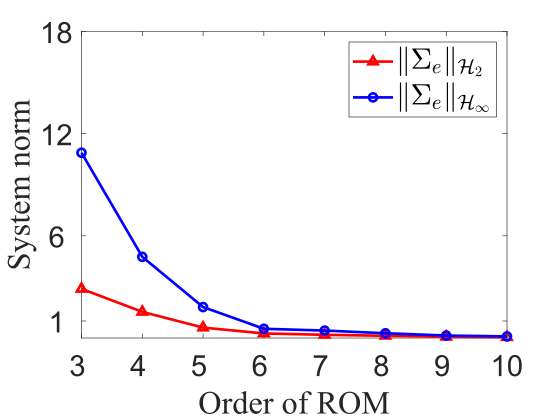}
} 
\hspace{1pt}
\subfloat[$(\alpha,\beta)=(1,1)$]{\label{fig:a1b1_error_sysnorm}
\includegraphics[width=0.31\textwidth]{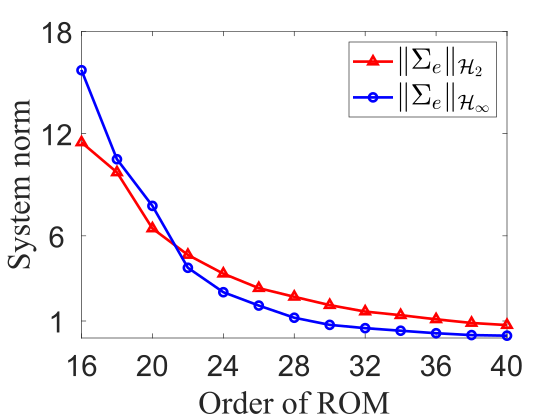}
}
\hspace{1pt}
\subfloat[$(\alpha,\beta)=(0,2)$]{\label{fig:a0b2_error_sysnorm}
\includegraphics[width=0.31\textwidth]{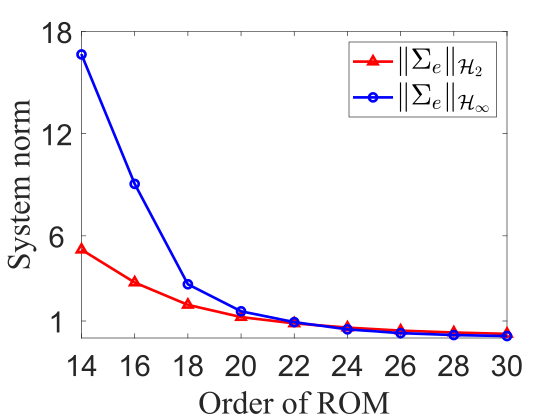}
}
\caption{In the balanced truncation approach, the error system norms are shown as a function of ROM order.} \label{fig:error_sysnorm}
\end{figure}

In conducting the BT approach for sensor selection, it is useful to evaluate system norms for the error dynamics associated with reduced-order balanced-truncation models as a function of model order $r_{BT}$.
In the present study, the actuator dynamics are non-trivial---they are modeled with integral effect---and so the balancing transformation and associated model reduction procedure are specifically conducted only with respect to the input-output response of the
fluid dynamic states $[\hat {\boldsymbol{v}}, \hat {\boldsymbol{\eta }}]^T$.
Both the $\mathcal{H}_2$- and $\mathcal{H}_{\infty}$-norm of the error system are reported in figure~\ref{fig:error_sysnorm}, for each of the three disturbances considered.
As expected, both system norms decrease with increasing model order.
In fact, it is possible to find $r_{BT}$ such that both system norms are less than a tolerance value $err$, where $err$ is small.
This indicates that the BT procedure will be able to identify a sparse sensor configuration for each disturbance by which SOF-LQR control will be able to recover full-information LQR control performance. 

In the ensuing sections, we will analyze the sensor selection results from the CE and BT procedures more closely.
We will do this for three types of disturbances: streamwise $(\alpha,\beta)=(1,0)$, oblique $(\alpha,\beta)=(1,1)$, and spanwise  $(\alpha,\beta)=(0,2)$.
In each case, the worst-case (TEG maximizing) disturbance associated with the specific control law will be considered.

\subsection{Streamwise disturbance~$(\alpha,\beta)=(1,0)$}
\begin{figure}
\begin{centering}
\subfloat[Maximum TEG ($G$) versus number of sensors.]{\label{fig:MTEG_compare_a1b0}
\includegraphics[width=0.9\textwidth]{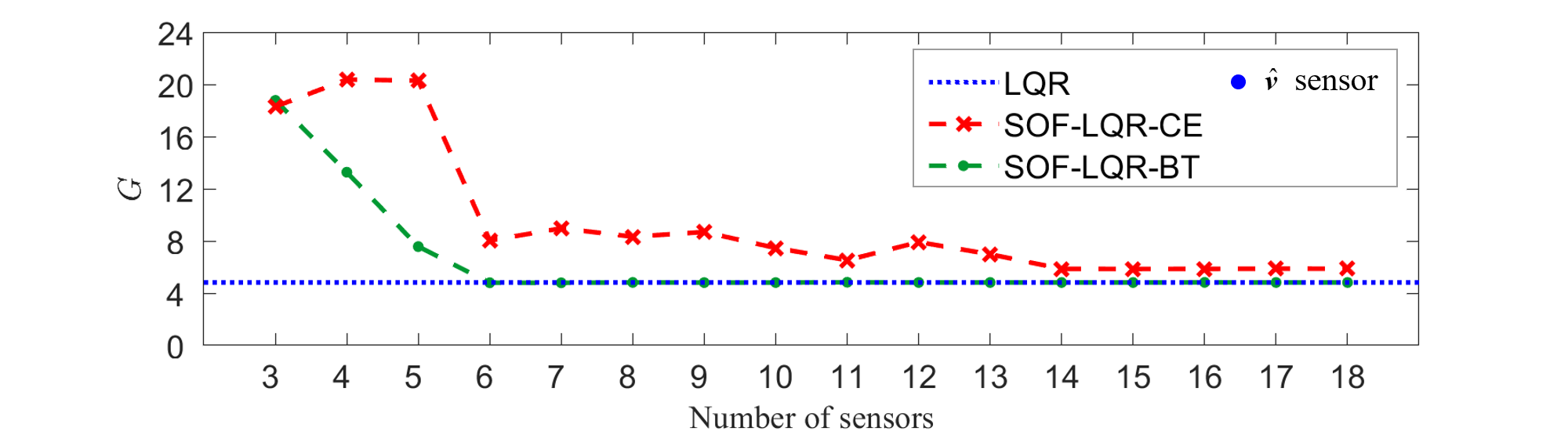} }
\vspace{1pt}
\subfloat[Sensor locations from CE approach.]{\label{fig:bycs_loc_tonum_a1b0}
\includegraphics[width=0.9\textwidth]{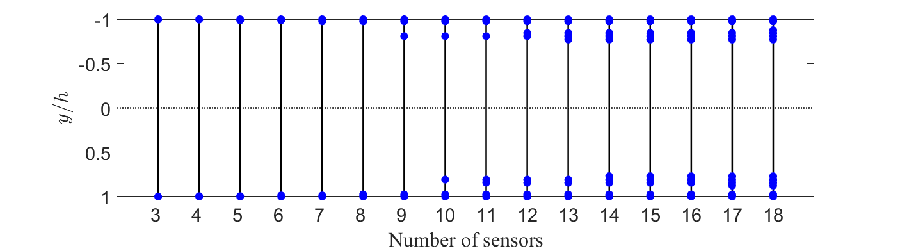} }
\vspace{1pt}
\subfloat[Sensor locations from BT approach.]{\label{fig:bybt_loc_tonum_a1b0}
\includegraphics[width=0.9\textwidth]{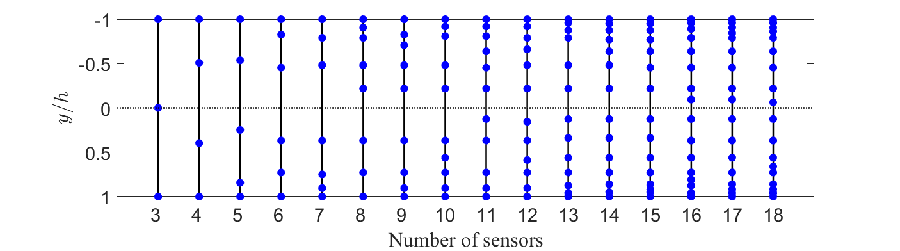} }
\caption{Controller performance and sensor configurations for worst-case streamwise disturbances with $(\alpha,\beta) = (1,0)$.} \label{fig:compare_2_a1b0}
\end{centering}
\end{figure}

The sensor configurations and the associated worst-case SOF-LQR control performance for streamwise disturbance with $(\alpha,\beta) = (1,0)$ are presented in figure~\ref{fig:compare_2_a1b0}. 
The maximum TEG of the full-information LQR controller~(blue dotted line), the maximum TEG~($G$) for the SOF-LQR controlled flow are reported as a function of the number of sensors in figure~\ref{fig:MTEG_compare_a1b0}, with CE configurations denoted as red crosses and BT configurations as green dots.
From this analysis, we see that SOF-LQR control recovers full-information LQR TEG performance when $r_{BT}\ge6$ sensors are used based on the BT approach.  Using the CE approach, SOF-LQR TEG performance recovers to within 5\% of the full-information LQR TEG when $r_{CE}\ge 14$ sensors are used.
In this case, the sufficiently rich library of candidate sensors~$C_n$ was constructed using only $\boldsymbol {\hat{v}}$ sensors.
As such, the observed differences between the sensor configurations from BT and CE is purely due to the locations of these sensors.
The specific sensor configurations obtained by the CE and BT sensor selection methods are illustrated in figure~\ref{fig:bycs_loc_tonum_a1b0} and figure~\ref{fig:bybt_loc_tonum_a1b0} respectively.
By both approaches, the near-wall sensors are identified as important for the control starting with as few as $r_{CE}=r_{BT}=3$ sensors.
Except for the case with $r_{CE}=9$, all the sensors selected by the CE approach are located either close-to-wall or clustered symmetrically about the channel centerline in the range $\vert y/h \vert = [0.77,0.84]$.
In contrast, the BT approach tends to yield asymmetrical arrangements in the channel and distributes sensors more uniformly between the channel walls.
That said, BT does tend to place a higher concentration of sensors in a similar range as that of the CE from $\vert y/h \vert = [0.79,1)$.
As we will see, near-wall $\boldsymbol {\hat{v}}$ information also tends to be important for performance recovery due to oblique and spanwise disturbances as well.

\begin{figure}
\captionsetup[subfigure]{justification=centering}
{
\includegraphics[width=0.9\textwidth]{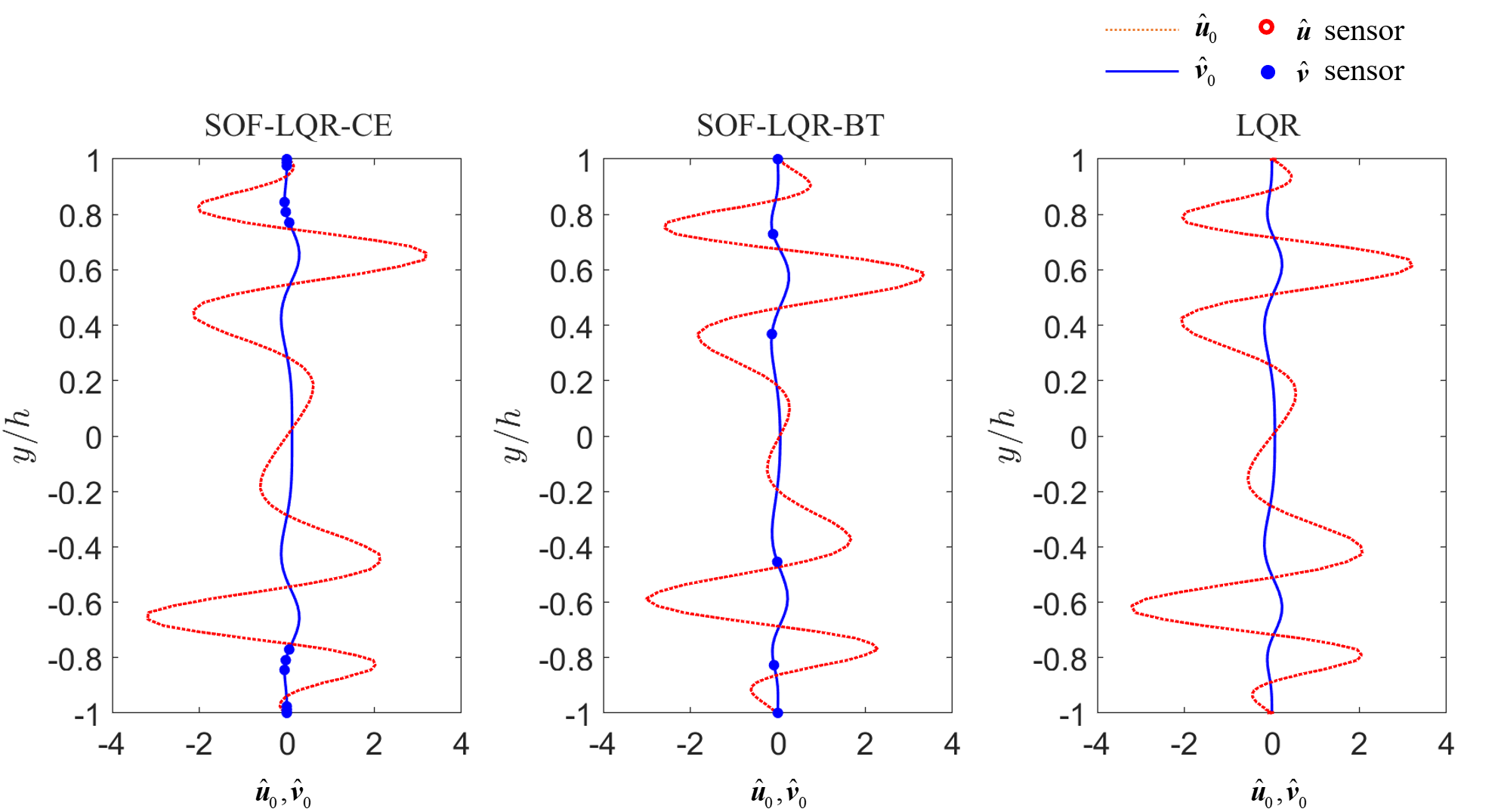}
}
\caption{The final selected sensor locations along with the optimal disturbance profile for wavenumber pair $(\alpha,\beta) = (1,0)$. Sensor number $r_{CE} = 14$, $r_{BT} = 6$.} \label{fig:vprofile_a1b0}
\end{figure}

We further investigate control performance for streamwise disturbances using the CE sensor configuration for $r_{CE}=14$ and the BT sensor configuration for $r_{BT}=6$.
The sensor locations along with the optimal disturbance profiles for the SOF-LQR controllers leading from the CE and BT sensor arrangements are shown in figure~\ref{fig:vprofile_a1b0}.
As can be seen here, the SOF-LQR controllers designed for sensor configurations from the CE and BT approaches yield similar optimal disturbance profiles as the full-information LQR controller.
The sensors selected by the CE approach capture the near-wall spatial features of the $\boldsymbol{\hat v}$ disturbance profile. The BT approach yields sensors arrangements that tend to capture spatial features throughout the channel.
Interesting, the largest qualitative differences in the SOF-LQR optimal disturbance profiles arise in $\boldsymbol {\hat{u}}$, even though the sensor configurations are restricted to only $\boldsymbol {\hat{v}}$ sensors.
 These differences---compared to the full-information case---are most visually prominent near the centerline for the BT configuration and near the walls for the CE configuration.

\subsection{Oblique disturbance~$(\alpha,\beta)=(1,1)$}
\begin{figure}
\begin{centering}
\subfloat[Maximum TEG ($G$) versus number of sensors.]{\label{fig:MTEG_compare_a1b1}
\includegraphics[width=0.9\textwidth]{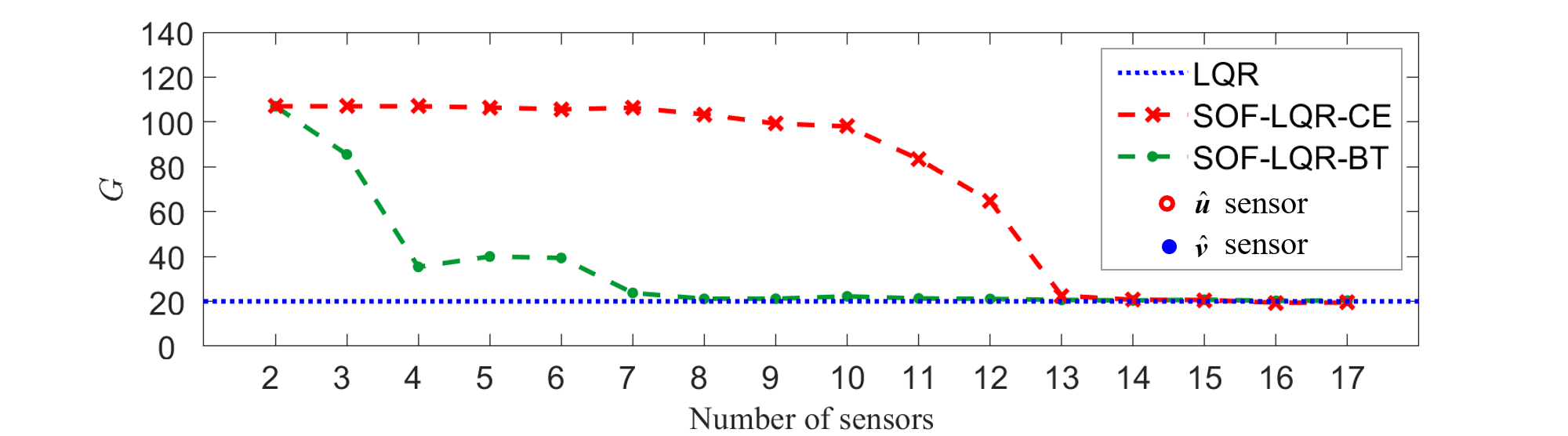} }
\vspace{1pt}
\subfloat[Sensor locations from CE approach.]{\label{fig:bycs_loc_tonum_a1b1}
\includegraphics[width=0.9\textwidth]{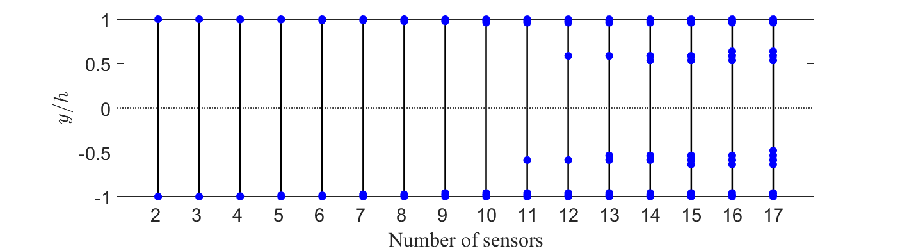} }
\vspace{1pt}
\subfloat[Sensor locations from BT approach.]{\label{fig:bybt_loc_tonum_a1b1}
\includegraphics[width=0.9\textwidth]{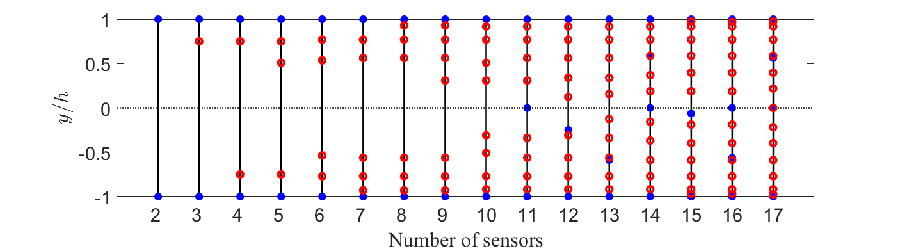} }
\caption{Controller performance and sensor configurations for worst-case oblique disturbances with $(\alpha,\beta) = (1,1)$.} \label{fig:compare_2_a1b1}
\end{centering}
\end{figure}

Sensor configurations and the associated worst-case SOF-LQR control performance for oblique disturbances with $(\alpha,\beta) = (1,1)$ are reported in figure~\ref{fig:compare_2_a1b1}.
Figure~\ref{fig:MTEG_compare_a1b1} illustrates the maximum TEG~($G$) performance as a function of the number of sensors associated with the sensor configurations determined by CE~(red cross) and BT~(green dots) methods relative to the full-information LQR control performance~(blue dotted line).
The CE and BT approaches both recover full-information LQR performance when at least $r_{CE} = 14$ and $r_{BT} = 13$ total sensors are used, respectively.
In figure~\ref{fig:bycs_loc_tonum_a1b1} and figure~\ref{fig:bybt_loc_tonum_a1b1}, the specific sensor configurations obtained by the CE and BT sensor selection methods are reported.
Note that both approaches were applied to the same candidate library of streamwise ($\boldsymbol {\hat u}$) and and wall-normal ($\boldsymbol {\hat v}$) sensors.
The BT method identifies both streamwise and wall-normal velocity sensors as important; whereas, the CE approach identifies that only wall-normal velocity sensors are important.
In both approaches, the near-wall wall-normal velocity sensors are immediately identified as important for control, starting with $r_{CE}=r_{BT}=2$ sensors.
As the number of sensors is increased, the CE approach identifies wall-normal velocity sensors in the range $\vert y/h \vert = [0.48,0.63]$ as important. 
In contrast, the BT approach does not select any $\boldsymbol {\hat v}$ sensors far from the walls until at least $r_{BT}\geq 11$ sensors are to be used.
Instead, the BT approach tends to place $\boldsymbol {\hat u}$ sensors at or near the centerline of the channel.
When $r_{BT} = 7$ or greater, SOF-LQR control based on the BT sensor arrangement yields TEG performance comparable to the full-information LQR controller.
This appears to be due to the placement of $\boldsymbol {\hat u}$ sensors in the range $\vert y/h \vert = [0.5,1)$.
The CE approach requires more sensors for this performance recovery.  
For SOF-LQR controllers based on the CE approach, TEG performance is comparable to the full-information LQR control when $r_{CE}\ge13$.
As in the BT case, the CE approach finds that information in the approximate range $\vert y/h \vert =[0.5,1)$ is important for recovering this performance, except now it is $\boldsymbol {\hat v}$ information that is deemed important.

\begin{figure}
\captionsetup[subfigure]{justification=centering}
{
\includegraphics[width=0.9\textwidth]{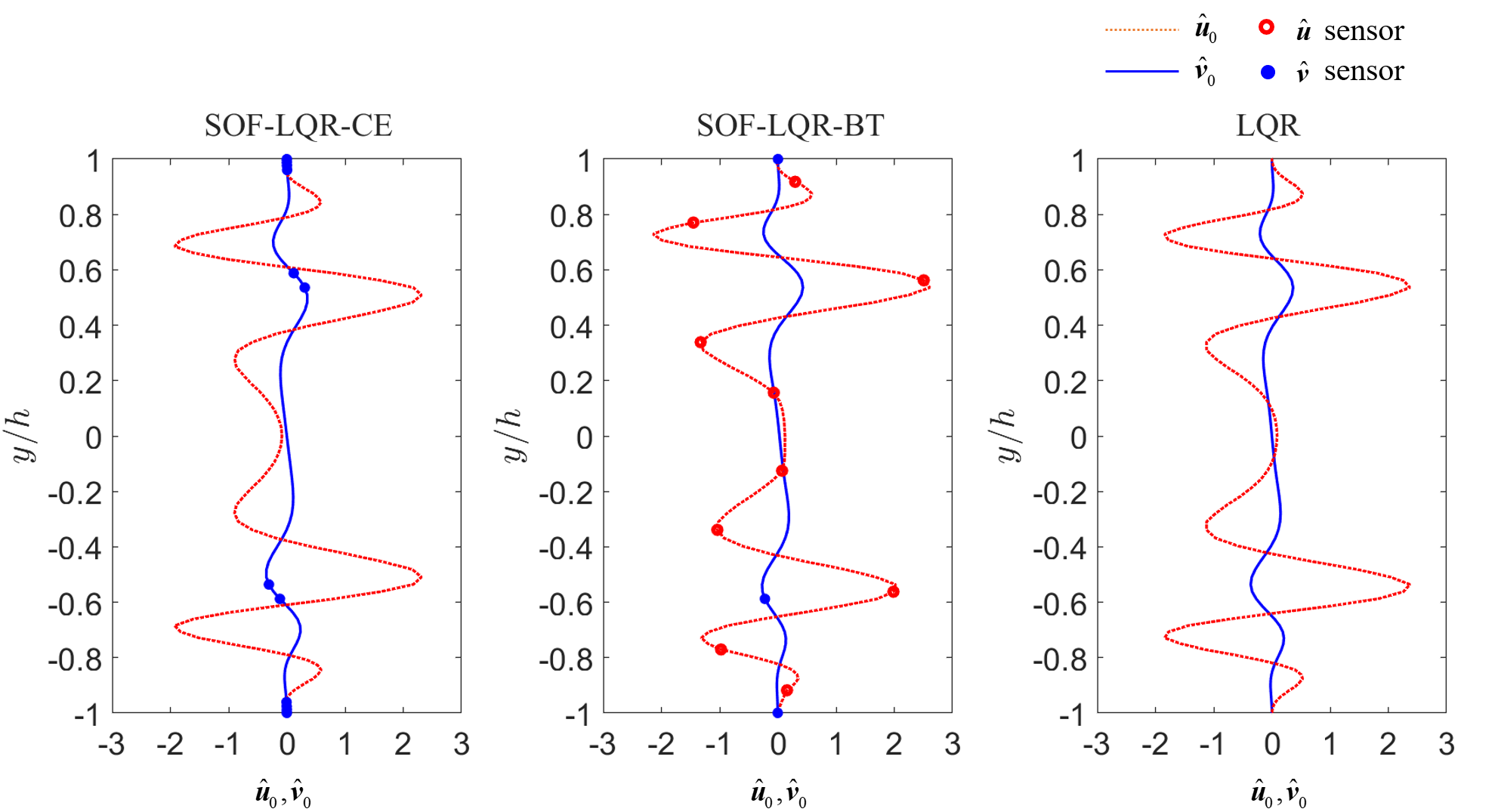}
}
\caption{The final selected sensor locations along with the optimal disturbance profile for wavenumber pair $(\alpha,\beta) = (1,1)$. Sensor number $r_{CE} = 14$, $r_{BT} = 13$.} \label{fig:vprofile_a1b1}
\end{figure}

We will investigate control performance for oblique disturbances in the remainder using the CE sensor configuration for $r_{CE} = 14$ and the BT sensor configuration for $r_{CE} = 13$.
The optimal disturbance profile leading to the maximum TEG for controllers based on these CE and BT sensor arrangements are reported in figure~\ref{fig:vprofile_a1b1}.
Similar to the streamwise disturbance case, the SOF-LQR optimal disturbance profiles from both the CE and BT sensor configurations are qualitatively similar to the full-information LQR optimal disturbance profiles.
In both the CE and BT approaches, the placement of $\boldsymbol {\hat v}$ sensors indicates that near-wall wall-normal velocity information is important.
In the CE case, the $\boldsymbol {\hat v}$ sensors capture the dominant spatial features of the $\boldsymbol {\hat v}$ disturbance profile. 
The distribution of $\boldsymbol {\hat v}$ sensors along the interior of the channel suggests that spatial derivatives $\boldsymbol {\hat v}$ are important.
In contrast, the BT approach introduces only a single $\boldsymbol {\hat v}$ sensor away from the walls, which captures information near the spatial peak in the optimal wall-normal velocity disturbance profile. 
The BT approach emphasizes sensors that capture the prominent spatial features of the $\boldsymbol {\hat u}$ optimal disturbance profile.

\subsection{Spanwise disturbance $(\alpha,\beta)=(0,2)$}

\begin{figure}
\begin{centering}
\subfloat[Maximum TEG ($G$) versus number of sensors.]{\label{fig:MTEG_compare_a0b2}
\includegraphics[width=0.9\textwidth]{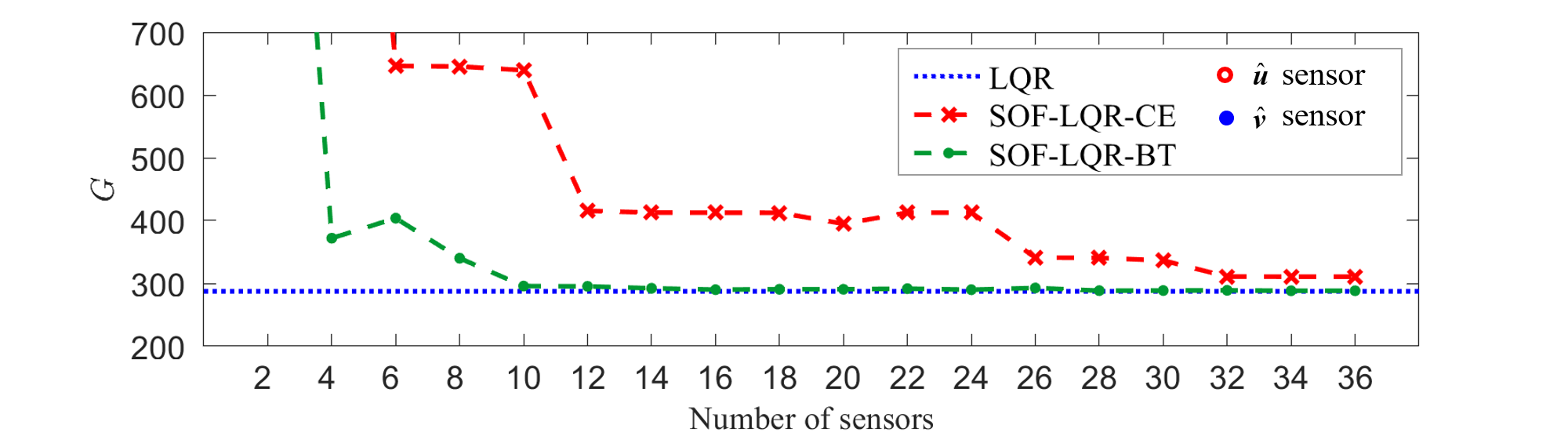} }
\vspace{1pt}
\subfloat[Sensor locations from CE approach.]{\label{fig:bycs_loc_tonum_a0b2}
\includegraphics[width=0.9\textwidth]{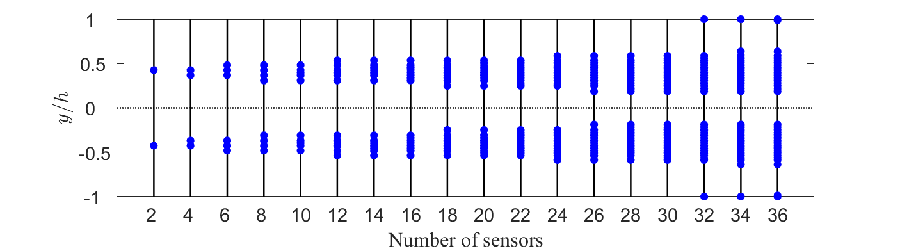} }
\vspace{1pt}
\subfloat[Sensor locations from BT approach.]{\label{fig:bybt_loc_tonum_a0b2}
\includegraphics[width=0.9\textwidth]{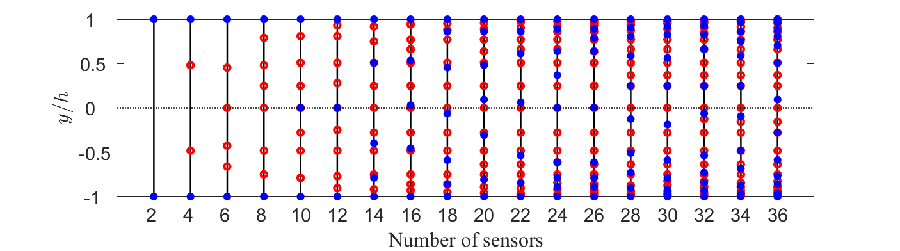} }
\caption{Controller performance and sensor configurations for worst-case spanwise disturbances with $(\alpha,\beta) = (0,2)$.} \label{fig:compare_2_a0b2}
\end{centering}
\end{figure}

Sensor configurations and the associated worst-case SOF-LQR control performance for spanwise disturbances with $(\alpha,\beta)=(0,2)$ are reported in figure~\ref{fig:compare_2_a0b2}.
The maximum TEG ($G$) for the controlled flow is reported as a function of the number of sensors for sensor configurations determined by CE (red crosses) and BT (green dots) methods in figure~\ref{fig:MTEG_compare_a0b2}.
These are compared with the maximum TEG~($G$) for the full-information LQR controller (blue dotted line).
From this analysis, it is evident that the BT approach yields a controller that recovers---at least approximately---the full-information control performance when $r_{BT}\ge10$. 
In contrast, the CE approach requires $r_{CE}\approx 32$ sensors to get within $1\%$ of the full-information TEG performance.
Investigating the specific sensor configurations obtained by the CE and BT sensor selection methods---see figures~\ref{fig:bycs_loc_tonum_a0b2} and \ref{fig:bybt_loc_tonum_a0b2}, respectively---provides some guidance on why this way be the case.
The BT method identifies both streamwise and wall-normal velocity sensors as important; whereas, the CE approach identifies that only wall-normal velocity sensors are important.
Even more significant, the BT approach immediately ($r_{BT}=2$) identifies that near-wall sensors are important for control.
In contrast, CE does not identify this same near-wall information as important until $r_{CE}\ge32$.
Up until $r_{CE}=32$, all of the sensors from the $CE$ approach are clustered symmetrically about the channel center line in the range $|y/h| = [0.187, 0.588]$.
The BT approach does not always yield a symmetric sensor arrangement, but the sensors are more evenly distributed throughout the interior of the channel.
Given the TEG performance achieved with these sensor configurations, it is evident that wall-normal information in the vicinity of the walls and at the channel center line are important for TEG reduction for spanwise disturbances.
Wall-normal velocity information at the center line is important, as the introduction of such a sensor allows the BT approach to recover full-information control performance with $r_{BT}=10$ sensors.
In some of the BT arrangements, the center line $\boldsymbol {\hat{v}}$ sensor is replaced by or augmented with a $\boldsymbol {\hat{u}}$ sensor.
In these cases, a pair of $\boldsymbol {\hat{v}}$ sensors tend to appear asymmetrically about the center line.

\begin{figure}
%\vspace{2pt}
\captionsetup[subfigure]{justification=centering}
{
\includegraphics[width=0.9\textwidth]{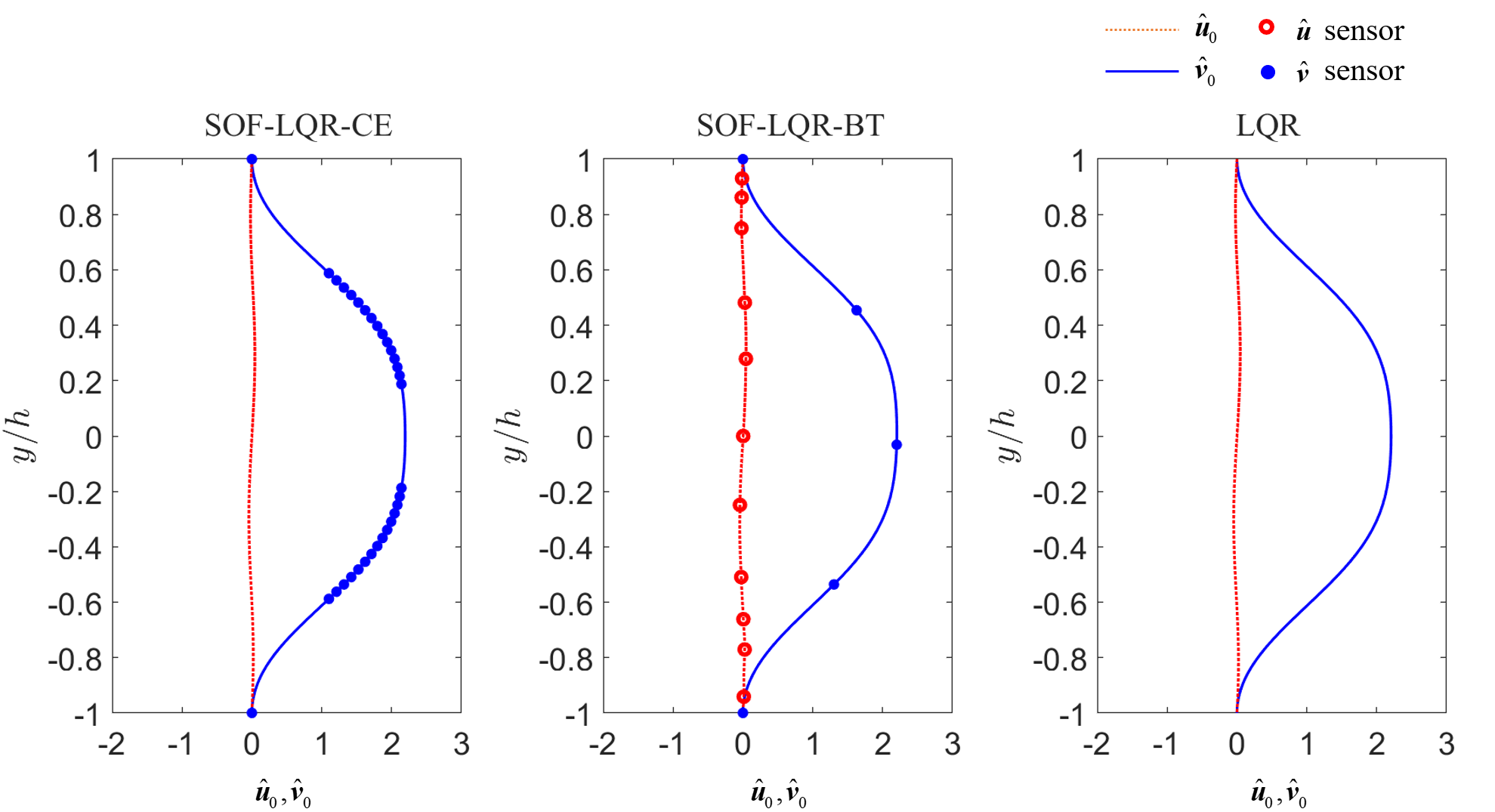}
}
\caption{The final selected sensor locations along with the optimal disturbance profile for wavenumber pair $(\alpha,\beta) = (0,2)$. Sensor number $r_{CE} = 32$, $r_{BT} = 16$.} \label{fig:vprofile_a0b2}
\end{figure}

We will investigate control performance for spanwise disturbances in the remainder using the CE sensor configuration for $r_{CE}=32$ and the BT sensor configuration for $r_{BT}=16$.
Note that the choice of $r_{BT}=16$ is motivated by the fact that the direct numerical simulations in section~\ref{sec:DNS} run to completion more quickly for this case than for $r_{BT}=10$.
The optimal disturbance profile leading to the maximum TEG for controllers based on these CE and BT sensor arrangements are reported in figure~\ref{fig:vprofile_a0b2}.
Interestingly, the optimal disturbance profiles for the SOF-LQR controllers are strikingly similar with one another and to optimal disturbance profile for the full-information LQR controller---even more so than in the previous cases considered for streamwise and oblique disturbances.
In some sense, this is to be expected based on previously reported findings regarding the $(\alpha,\beta)=(0,2)$ disturbance case.
This wave-number pair exhibits the highest TEG among all wave-number pairs, and is the least affected by control~\citep{Kalur2019,Martinelli2011,sun2019}.
In both the CE and BT approaches, we see that the $\boldsymbol {\hat{v}}$ sensors capture the dominant spatial features of the $\boldsymbol {\hat{v}}$ disturbance profile.
However, only the BT sensor placements are able to capture the spatial profile associated with the $\boldsymbol {\hat{u}}$ component of the optimal disturbance.
Although the $\boldsymbol {\hat{u}}$ component is smaller in magnitude compared with the $\boldsymbol {\hat{v}}$ component, this has an important consequence for TEG control performance.

\subsection{Controller robustness analysis} \label{sec:robustness}

Up to this point, all of the control performance analysis has been conducted for nominal conditions at $Re=3000$.
In this section, we will investigate the robustness of the TEG reduction when the controllers designed for $Re=3000$ are applied at ``off-design'' sub-crital $Re$ values.
The specific sensor arrangements we will investigate here correspond to the cases with a corresponding detailed analysis in the sections above.
These are summarized in table~\ref{tab:final_case}.
\begin{table}
    \centering
    \begin{tabular}{m{1.2cm}m{4.5cm}m{4.5cm}}
         
         $(\alpha,\beta)$   &   Balanced Truncation & Gain Column Norm Evaluation  \\ \hline
         
        {(1,0)}              & 6 $v$-sensors    & 14 $v$-sensors \\
         ~\\
         (1,1) & 10 $u$-sensors    & 0 $u$-sensors \\
                             & 3 $v$-sensors    & 14 $v$-sensors \\
                             ~\\
         {(0,2)}  & 11 $u$-sensors    & 0 $v$-sensors \\ 
                      & 5 $v$-sensors    & 32 $v$-sensors \\
                      ~\\
    \end{tabular}
    \caption{Sensor configurations used for detailed performance evaluation.}
    \label{tab:final_case}
\end{table}

\begin{figure}
\captionsetup[subfigure]{justification=centering}
\subfloat[$(\alpha,\beta)=(1,0)$ $r_{CE}=14$, $r_{BT}=6$][$(\alpha,\beta)=(1,0)$ \\ $r_{CE}=14$, $r_{BT}=6$]{\label{fig:a1b0_robustness}
\includegraphics[width=0.31\textwidth]{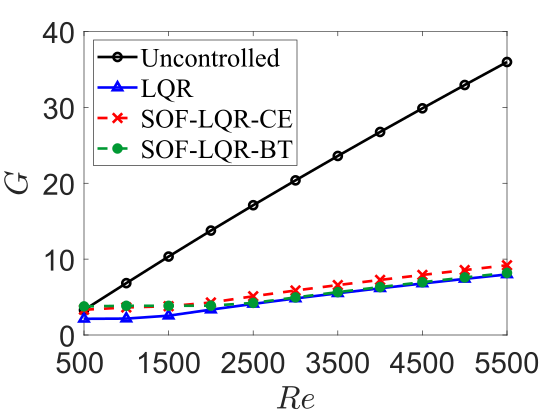}
} 
\hspace{1pt}
\subfloat[$(\alpha,\beta)=(1,1)$ $r_{CE}=14$, $r_{BT}=13$][$(\alpha,\beta)=(1,1)$ \\ $r_{CE}=14$, $r_{BT}=13$]{\label{fig:a1b1_robustness}
\includegraphics[width=0.31\textwidth]{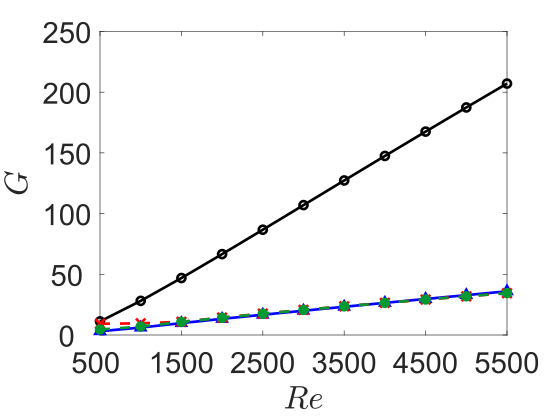}
}
\hspace{1pt}
\subfloat[$(\alpha,\beta)=(0,2)$ $r_{CE}=32$, $r_{BT}=16$][$(\alpha,\beta)=(0,2)$ \\ $r_{CE}=32$, $r_{BT}=16$]{\label{fig:a0b2_robustness}
\includegraphics[width=0.31\textwidth]{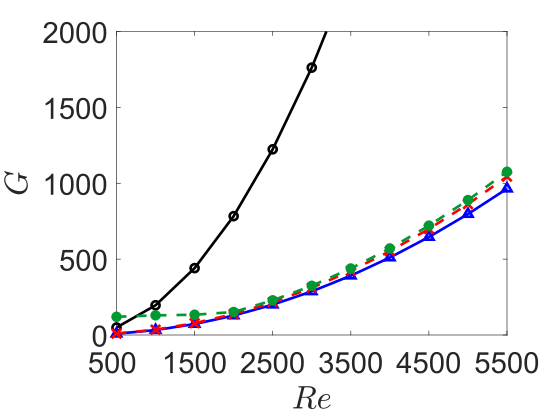}
}
\caption{Robustness analysis} \label{fig:robustness}
\end{figure}

To investigate robust TEG performance, we again consider
closed-loop system responses for optimal perturbations that result in the maximum TEG~($G$) for a given closed-loop system with control applied at the off-design $Re$.
Results for SOF-LQR controllers designed at $Re=3000$ then applied at $Re=[500,5500]$ are reported in figure~\ref{fig:robustness}.
The performance of both controllers at off-design Reynolds numbers is strikingly similar to the corresponding performance at on-design conditions at Reynolds numbers higher than $1000$.
These results demonstrate that SOF-LQR controllers can be designed based on BT and CE sensor arrangements \emph{robustly} recover full-information LQR control performance;
however, this robustness is not unconditional.

For streamwise wave disturbances at $Re\le500$, the SOF-LQR controllers designed for $Re=3000$---using either CE or BT sensor arrangements---fail to reduce TEG relative to the uncontrolled flow. 
For spanwise wave disturbances at $Re\le1000$, the SOF-LQR designed for $Re=3000$ based on the BT sensor arrangement lacks the same robust performance recovery characteristics as the CE sensor arrangement.
In fact, the BT arrangement fails fails to reduce TEG relative to the uncontrolled flow at $Re\le500$.
However, designs based on the CE sensor arrangement maintain the TEG reduction even at these lower Reynolds numbers.
As the TEG under smaller Reynolds number is relatively low, it is not surprised that the TEG reduction ability also decreases. 
Though there are less chances that the transition will happen compare to higher Reynolds numbers, this suggests us that we need to design the SOF-LQR controllers separately for TEG reduction at low Reynolds numbers.
Indeed, SOF-LQR controllers designed using either BT and CE sensor arrangements at ``on-design'' conditions would recover the full-information LQR performance and reduce TEG relative to the uncontrolled flow.

\section{Results: nonlinear performance analysis using direct numerical simulations} \label{sec:DNS}

Sensor selection for controller performance recovery was investigated using linear analysis in the previous section.
Here, we are interested in determining the reliability of the resulting sensor-based output feedback control laws for transient energy growth reduction and transition control in the nonlinear setting.
To this end, three-dimensional direct numerical simulations of plane Poiseuille flow at $Re=3000$ are performed using a modified version of the spectral code {\it Channelflow} \citep{GibsonHalcrowCvitanovicJFM08,channelflow} to solve the incompressible Navier--Stokes equations. A second-order semi-implicit Crank-Nicolson Runge--Kutta temporal scheme is used.   
In past studies, the domain size is usually set to fit one wavelength of the wavenumber of interest for each direction. However, as the transition includes large-scale deformation of the original perturbation structures, the underlying physics during the transition processes cannot be fully captured with a limited domain size. Hence, we use a rectangular computational domain of size $8\pi h \times 2h \times 2\pi h$ in $x$-, $y$- and $z$-directions, respectively, which is relatively large in space to capture a broader extent of flow physics involved in the transition process. This provides a more realistic setting to assess transition control performance.  To discretize the flow field, $N=101$ Chebyshev points are specified in the $y$-direction, and $128\times64$ points are uniformly spaced along $x$- and $z$- directions, respectively. Periodic boundary conditions are assumed in the $x$- and $z$-directions in which the flow variables (velocity and pressure) are represented by Fourier expansion. No-slip boundary condition is specified at upper and lower walls for the uncontrolled flow. In the controlled flow, wall-normal velocity $\boldsymbol v(\pm h)$ is determined via feedback control law, while streamwise and spanwise velocities are zeros at the walls. Moreover, a constraint of constant bulk velocity is specified. For both baseline and controlled flows, grid resolution studies with doubled grids in each direction have been performed to ensure accuracy of results. 

Actuation in the form of blowing and suction in $y$-direction is implemented on the entire upper and lower walls via temporally changing amplitude of wall-normal velocity $\boldsymbol{v}(\pm h)$. Because the control strategy is designed based on the linear dynamical system discussed above, both actuation and measurement are associated with the same wavenumber pair $(\alpha,\beta)$ as the optimal perturbation, but implemented within the nonlinear simulations; 
this allows flow control mechanisms to be isolated and investigated. In practice, a bank of linear controllers over all wave number combinations would be most effective.
The value of wall-normal velocity on walls is determined by the feedback control law described above in section \ref{sec:sensor_selection}.
 
In the linear analysis of section \ref{sec:results}, characteristics of the optimal disturbance were examined solely using the linear dynamics. In the direction numerical simulations, baseline and controlled flows have an initial condition consisting of the base flow $[\boldsymbol{\bar u}, \boldsymbol{\bar v}, \boldsymbol{\bar w}]=[1-(y/h)^2,0,0]$ and optimal disturbance $[\boldsymbol{u}'_0,\boldsymbol{v}'_0,\boldsymbol{w}'_0]$ with a small amplitude. The kinetic energy density of the initial optimal disturbance is denoted by $E_0$. Moreover, a random perturbation is also introduced with kinetic energy density of $1\%$ of $E_0$ into the flow to expedite an emergence of a laminar-to-turbulent transition in the flows, but has negligible influence on the feature of the optimal disturbance \citep{Reddy:JFM98}.

\subsection{Laminar-to-turbulent transition} \label{sec:full_state_lqr}

\begin{figure}
\subfloat[$(\alpha,\beta)=(1,0)$]{\label{fig:a1b0_op_lqr}
\includegraphics[width=0.27\textwidth]{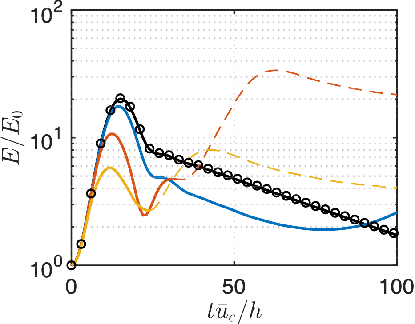}
} 
\hspace{1pt}
\subfloat[$(\alpha,\beta)=(1,1)$]{\label{fig:a1b1_op_lqr}
\includegraphics[width=0.27\textwidth]{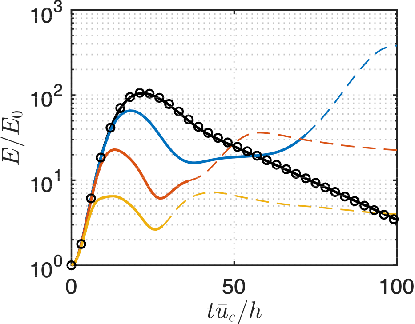}
}
\hspace{1pt}
\subfloat[$(\alpha,\beta)=(0,2)$]{\label{fig:a0b2_op_lqr}
\includegraphics[width=0.395\textwidth]{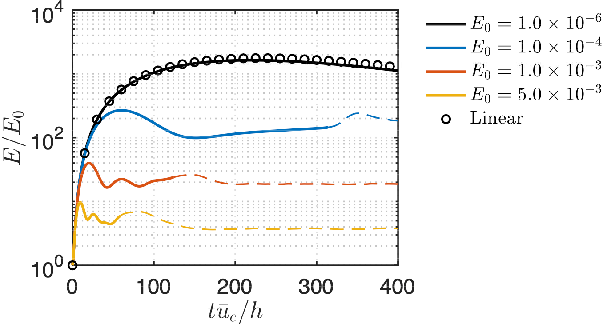}
}
\caption{Transient energy growth of optimal disturbance with amplitude $E_0$. Dashed lines indicate that flow is turbulent state.} \label{fig:Base_TEG}
\end{figure}

Direct numerical simulations are performed with various optimal disturbance amplitudes for each wavenumber pair. As shown in figure \ref{fig:Base_TEG}, the transient energy growth density for the smallest disturbance amplitude considered overlaps the linear result. Moreover, we find that the amplification of initial disturbance is suppressed as the amplitude of disturbance increases and nonlinear effects become prominent. Although smaller transient energy growth is observed in the cases with larger initial disturbances, the laminar-to-turbulent transition still emerges as indicated by dashed lines in figure \ref{fig:Base_TEG}. Indeed, large values of the absolute transient energy $E$, rather than the amplification $E/E_0$, is more likely to lead to transition. This also indicates that for the same initial disturbance, an ability to reduce the transient amplification can serve to suppress the laminar-to-turbulent transition. 

\subsection{Transition suppression}

In the controlled flow, the sensor selection strategy evaluated in the direct numerical simulations are listed in table \ref{tab:final_case}. These strategies require the least number of sensors in output feedback control to recover the full-state LQR control performance in terms of reducing transient energy growth. For the streamwise disturbance with $(\alpha,\beta)=(1,0)$, the threshold of minimal seed that leads to a laminar-to-turbulent transition increases from $E_0=1.0\times 10^{-4}$ to $E_0=1.0\times 10^{-3}$. For the oblique disturbance, the threshold increases from $E_0=5.0\times 10^{-5}$ to $E_0=5.0\times 10^{-4}$. However, for the spanwise disturbance, the controllers do not suppress transition, and actually lead to an earlier emergence of transition. This occurs for the full-information controller as well \citep{sun2019}. Although TEG performance is recovered, this case shows that TEG reduction is not the only objective to be considered when it comes to transition control.   

\subsubsection{Oblique and streamwise disturbances}

Because the transition and control mechanisms in oblique $(\alpha,\beta)=(1,1)$ and streamwise $(\alpha,\beta)=(1,0)$ disturbances are similar, we will mainly use oblique case as an example to discuss how the controller modifies the flow and ultimately suppresses laminar-to-turbulent transition.

\begin{figure}
    \centering
    \includegraphics[width=1.0\textwidth]{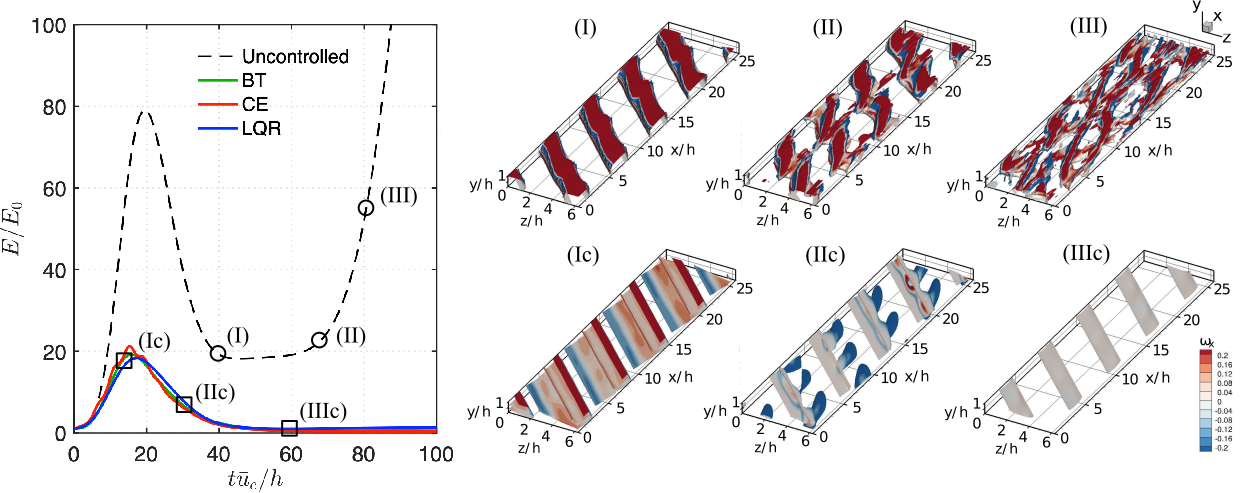}
    \caption{Transient energy growth energy of uncontrolled flow and controlled flows ($E_0 =5\times 10^{-5}$) with controllers designed from LQR, balanced truncation, and gain column norm evaluation methods. Inserts are corresponding iso-surface of $Q$-criterion \citep{Hunt:88} colored by streamwise vorticity $\omega_x$.}
    \label{fig:DNS_flowfield}
\end{figure}

In the case of oblique disturbance with optimal disturbance amplitude of $E_0=5\times 10^{-5}$, the transient energy growth density reaches maximum amplification of $E/E_0\approx 80$ at $t\boldsymbol{\bar u}_c/h=20$ as shown in figure \ref{fig:DNS_flowfield}. Representative flowfields at different stages are displayed to illustrate the transition mechanism. Before stage (I), oblique coherent structures develop from small to large scale as transient energy grows. At stage (I), although the transient increase of perturbation kinetic energy has decreased to a relatively lower value, a large value of streamwise vorticity is observed along the oblique coherent structures near the wall. At stage (II), streamwise vortical structures start to grow stemming from the oblique coherent structures. As these structures develop, they interact with each other and break down into small scale structures. From this, a laminar-to-turbulent transition emerges at stage (III).    In all the three controlled flows, the maximum transient energy growth densities are reduced to $E/E_0\approx 20$ around $t\boldsymbol{\bar u}_c/h=18$, and their trajectories are similar. 
The control mechanism is similar among the SOF-LQR controllers based on BT and CE sensor configurations and the full-information LQR controller. Here we use snapshots of flow fields with the SOF-LQR controller designed using the sensor configuration from the BT method as an example in figure \ref{fig:DNS_flowfield}.
At stage (Ic) in the controlled flow, the coherent structures remain almost uniform in the oblique direction, with smaller values of streamwise voriticty generated near the walls. As transient energy decreases at stage (IIc), the only oblique structures observed in figure \ref{fig:DNS_flowfield} are induced from the wall actuation. As time further elapses, the streamwise vortical structures decay and actuation amplitude also decreases, ultimately suppressing the transition observed in the uncontrolled flow. 

\begin{figure}
    \centering
    \includegraphics[width=0.9\textwidth]{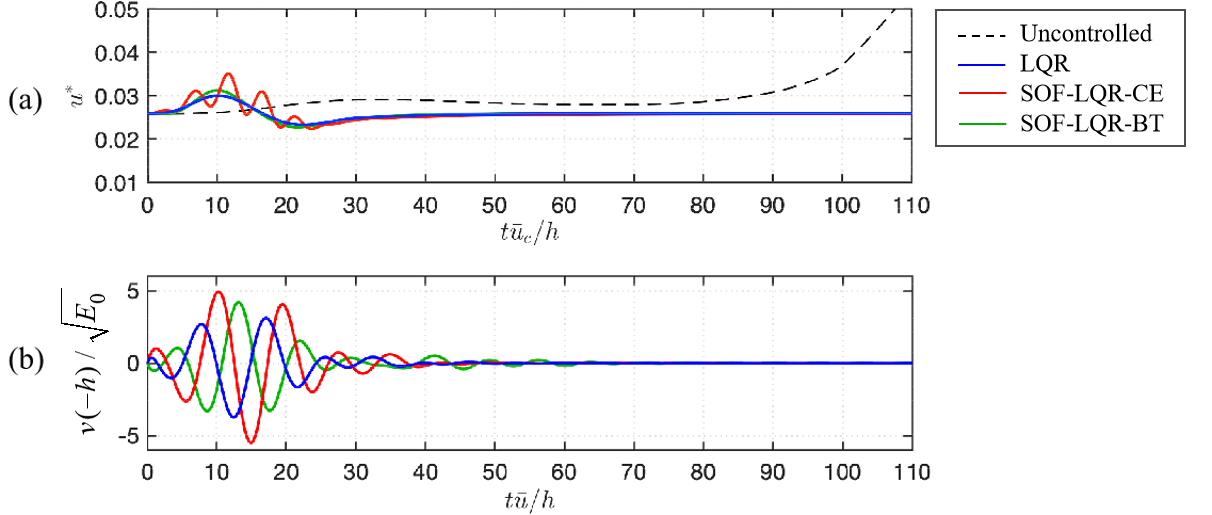}
    \caption{(a) Friction velocity and (b) Fourier coefficient of wall-normal velocity boundary condition $v(-h)$ at lower wall of flow ($E_0= 5\times 10^{-5}$) with $(\alpha,\beta)=(1,1)$.}
    \label{fig:A1B1_friction}
\end{figure}

Friction velocity has been used as a quantity to identify the laminar-to-turbulent transition in the flow, which is defined as
\begin{equation}
    u^*=\sqrt{\frac{\tau_w}{\rho}},
\end{equation}
where $\tau_w=\mu(\partial u/\partial y)$, $\mu$ is dynamic viscosity, and the density $\rho$ is set to one for the incompressible flow. The friction velocity at the lower wall is calculated using mean flow quantity. The time-history of the friction velocity for uncontrolled and controlled flows are shown in figure \ref{fig:A1B1_friction}. A sudden increase in friction velocity in the uncontrolled flow (see figure \ref{fig:A1B1_friction} (a)) indicates an emergence of the transition. In the controlled flows, the actuation is turned on to introduce blowing and suction during the transient energy growth period (see figure \ref{fig:A1B1_friction}). Correspondingly, we also observed that the friction velocity varies as the actuation is active. 
As the kinetic energy grows, the friction velocity increases. When the kinetic energy decreases, the friction velocity reduces to a lower value and then stays at a constant value. Particularly, the friction velocity with the controller designed using the CE sensor configuration has a high-frequency oscillation during the transient process, but the general trend is similar to the other two controlled cases.

\begin{figure}
    \centering
    \includegraphics[width=0.6\textwidth]{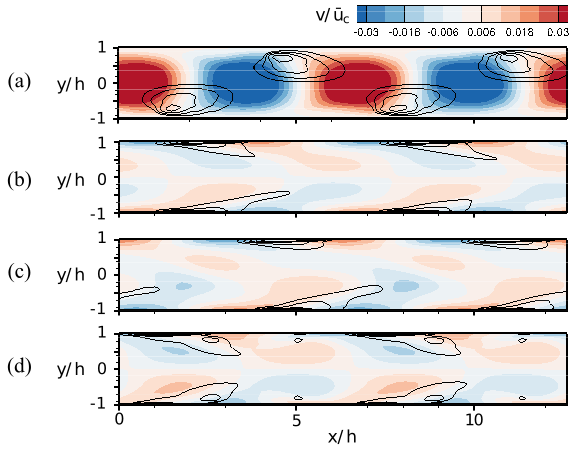}
    \caption{Contours of wall-normal velocity and $Q$-criterion over a range of [0.01, 0.05] at slice of $z/h=0$ of flow ($E_0= 5\times 10^{-5}$) with $(\alpha,\beta)=(1,1)$ at $t\bar u_c/h=18$ (a) Uncontrolled flow, and controlled flow with controller designed by (b) LQR, (c) balanced truncation, and (d) gain column norm evaluation methods.}
    \label{fig:v_Q}
\end{figure}

Slices of uncontrolled and controlled flowfields are displayed in figure \ref{fig:v_Q} to show modifications of the flow features due to wall actuation. The slices are extracted at convective time of $t\boldsymbol{\bar u}_c/h=18$, when the kinetic energy density reaches its maximum value. In the uncontrolled flow, large wall-normal velocity fluctuations reside in the center between the upper and lower walls. The size of the coherent structures highlighted by $Q$-criterion is about half height of the channel as the transient energy grows. From the flowfields shown in figure \ref{fig:DNS_flowfield} and our previous study \citep{sun2019}, the existence of these large coherent structures induces a large value of streamise vorticity and generates streamwise vortical structures. Once these vortical structures break down, a transition to turbulence emerges. In all the three controlled flows, we observed that the wall-normal velocity actuation modifies the pattern of velocity flowfields and changes the distribution of high-shear regions. This change of flow hinders the formation of large coherent structure, which in turn prevents laminar-to-turbulent transition from arising in the flow.    

\subsubsection{Spanwise disturbance}

\begin{figure}
    \centering
    \includegraphics[width=0.9\textwidth]{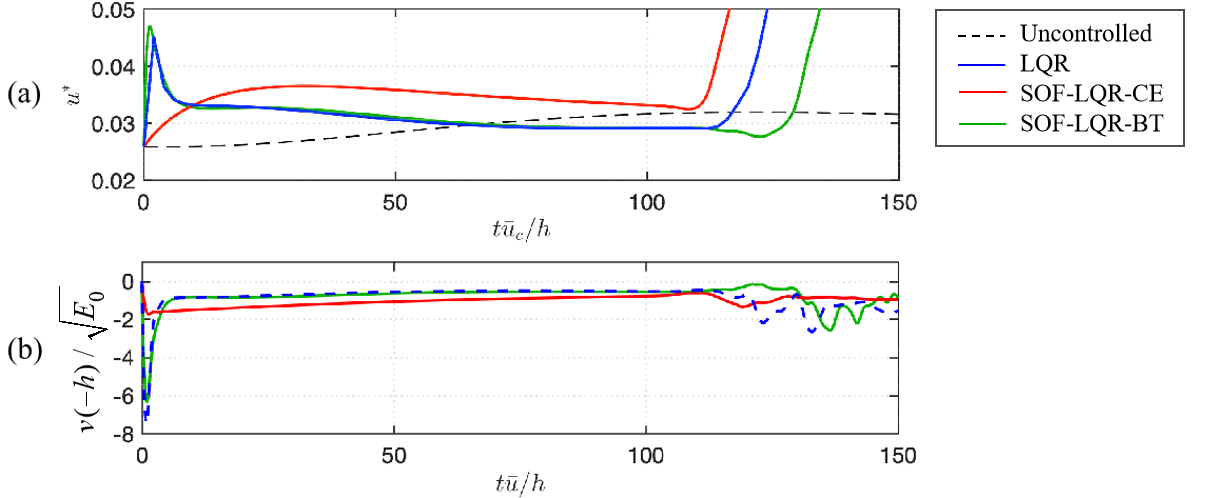}
    \caption{(a) Friction velocity and (b) Fourier coefficient of wall-normal velocity boundary condition $v(-h)$ at lower wall of flow ($E_0= 5\times 10^{-5}$) with $(\alpha,\beta)=(0,2)$.}
    \label{fig:A0B2_friction}
\end{figure}

In the cases with spanwise $(\alpha,\beta)=(0,2)$ optimal disturbance considered in the present work, none of the controllers increase the threshold of laminar-to-turbulent transition. In fact, the controllers actually worsen the transition scenario. As shown in figure \ref{fig:A0B2_friction} (a) of case with $E_0=5\times10^{-5}$, there is no laminar-to-turbulent transition in the uncontrolled flow, but the controllers trigger the transition. While the controller is active, actuation amount in the form of blowing and suction gradually decreases as shown in figure \ref{fig:A0B2_friction} (b). As being closely examined in our previous study \citep{sun2019}, the control input introduces streamwise vortices near walls, which hinder the merging process of vortex pairs in the center of the channel, resulting in a reduction of transient energy growth. However, the control input creates high-shear regions near the wall where secondary instabilities creep in and cause a laminar-to-turbulent transition. This phenomenon suggests that decreasing transient energy growth from linear analysis is not sufficient to suppress (or even delay) the transition. 

\begin{figure}
    \centering
    \includegraphics[width=0.9\textwidth]{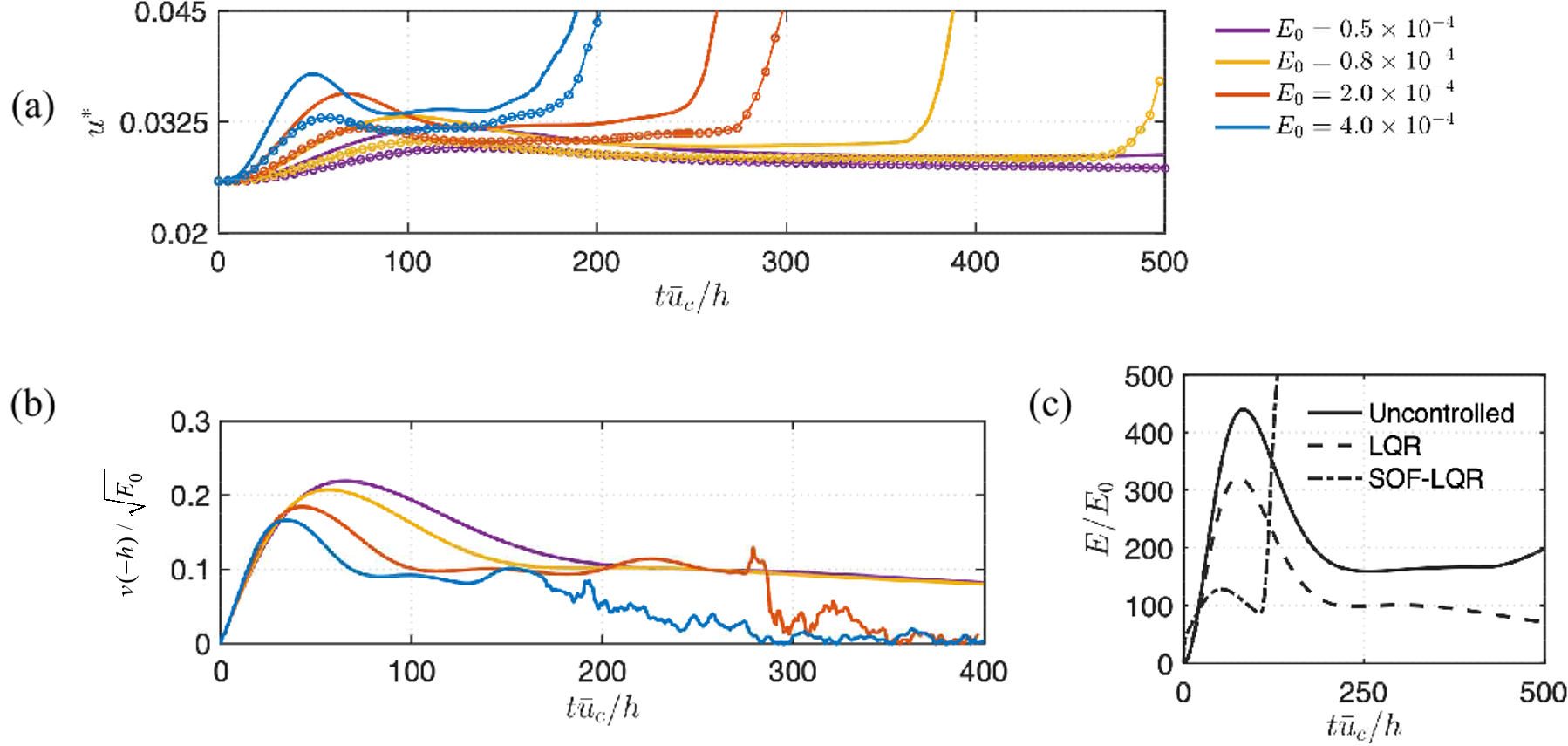}
    \caption{With only shear-stress sensors at the walls, (a) friction velocity of uncontrolled and controlled cases. Solid line with circles represents controlled case. (b) Fourier coefficient of wall-normal velocity boundary condition $v(-h)$ in controlled flows. (c) Transient energy growth of uncontrolled and controlled flow with disturbance amplitude of $E_0=5\times 10^{-5}$. }
    \label{fig:A0B2_friction_SOF}
\end{figure}

Interestingly, using the same objective function, but limiting the information available for static output feedback control to wall shear stress measurements only, actually enables transition suppression in this case, and transition delay for large amplitude disturbances \citep{Sun:APS19} as shown in figure \ref{fig:A0B2_friction_SOF} (a). During the time period considered in the present study, the SOF-LQR controller does not worsen the transition and leads to a lower friction velocity compared to uncontrolled flow. As seen in figure \ref{fig:A0B2_friction_SOF} (b), the normalized actuation velocity is smaller than the one shown in figure \ref{fig:A0B2_friction}. 
Correspondingly, the reduction of transient energy growth with SOF-LQR control with only shear stress measurements is roughly 40\% of the reduction achieved by full-information LQR with respect to maximum transient energy growth (see figure \ref{fig:A0B2_friction_SOF} (c)). This SOF-LQR with shear-stress sensors at the walls is able to modify the flow sufficiently to delay transition and does so with a gentle actuation such that the control does not trigger other instabilities in the nonlinear flow that cause a transition. 

\section{Discussion and conclusions} \label{sec:conclusions}
  
In this paper, we proposed and investigated two approaches for sensor selection to enable sensor-based output feedback controllers to recover full-information performance.
Both approaches leveraged the fact that, with a sufficiently rich set of sensors, a static output feedback~(SOF) controller can always be designed to recover full-information control performance exactly.
Thus, we showed that the problem of sensor selection for performance recovery could be recast as a problem of sensor down-selection from a sufficiently rich library of candidate sensors.
One approach for sensor selection was based on a column-norm evaluation~(CE) of the SOF controller gain matrix used to achieve full-information performance; the other approach was based on a balanced truncation~(BT) procedure for model order reduction of linear dynamic systems.

Both sensor-selection methods were demonstrated on the problem of transient energy growth~(TEG) reduction and transition suppression in a channel flow at the sub-critical Reynolds number of $Re=3000$ using linear quadratic optimal control strategies.
All sensor-based output feedback controllers in this study were designed based on SOF-LQR synthesis and compared with full-information LQR controllers designed based on the same quadratic control objective.
Based on linear performance analyses, both the CE and BT sensor selection approaches identified sensor configurations that enabled SOF-LQR control to recover full-information LQR control performance for optimal streamwise, oblique, and spanwise disturbances.
In all cases where performance recovery was achieved, the optimal disturbance profiles for the SOF-LQR controlled flows were qualitatively similar to those for the full-information LQR controlled flows.
These similarities were most striking in the case of spanwise disturbances, but all corroborate the fact that the dynamic responses of of all of these controlled systems are approximately the same.
Sensors selected by the BT and CE approaches tended to be placed in the vicinity of prominent spatial features associated with these optimal disturbance profiles.
Further, we found that SOF-LQR controllers designed based on these sensor configurations also exhibited robustness to variations in the Reynolds number.
Controllers designed for $Re=3000$ continued to achieve comparable worst-case TEG performance to the full-information LQR controller when applied over a range of ``off-design'' sub-critical Reynolds numbers.

In general, the BT approach required fewer sensors to recover full-information control performance compared to the CE approach.
Wall-normal velocity information near the channel walls was consistently determined to be important for control performance by both methods over all disturbances considered.
However, the arrangement and type of other sensors between the CE and BT approaches differed otherwise.
Our results for the oblique and spanwise disturbance designs suggest that TEG control benefits from streamwise velocity information---more so than wall-normal velocity information---along the channel interior.
For these cases, both streamwise and wall-normal velocity sensors were available in the candidate sensor library.  Yet, the CE approach consistently removed streamwise velocity sensors in favor of wall-normal velocity sensors, which may contribute to the need for more sensors for performance recovery relative to the BT approach. 
The BT approach consistently yielded a heterogeneous set of sensors, with wall-normal velocity sensors near the walls and streamwise velocity sensors along the channel interior.

In addition to the linear analysis, we conducted direct numerical simulations~(DNS) to evaluate control performance in the nonlinear setting. 
SOF-LQR controllers designed based on both sensor selection approaches continued to recover full-information controller performance in DNS.
For streamwise and oblique disturbances, the SOF-LQR controllers successfully increased the disturbance energy threshold for transition by an order of magnitude.
We found that control hindered the formation of large coherent structures, in part because actuation from the wall served to modify the shear distribution in the flow. 
With a reduced size of coherent structures, the amount of induced streamwise vortical streaks decreased and ultimately suppressed the transition that was observed in the uncontrolled flow. 
We also note that TEG reduction observed in linear analysis was not sufficient for preventing transition in the nonlinear simulations. 
The linear analysis missed the secondary instabilities that were excited by actuation in the case of spanwise disturbances. 
These nonlinear interactions promoted an earlier transition, even though the TEG was reduced.  
Controller designs that explicitly account for these nonlinear interactions should be investigated in future studies.

Finally, we note that the sensor selection methods presented here are more generally applicable to other performance objectives, sensor-types, and flow configurations.
In addition, both the CE and BT approaches introduced for sensor selection in this study can be trivially extended to the problem of actuator selection, simply by considering the dual problem and a sufficiently rich library of candidate actuators.
We hope that the promising results demonstrated in this paper will lead to further adoption and refinement of these strategies into the future.

\section{Acknowledgments}
This material is based upon work supported by the Air Force Office of Scientific Research under award number FA9550-19-1-0034, monitored by Dr. Gregg Abate, and the National Science Foundation under award number CBET-1943988, monitored by Dr. Ronald D. Joslin.

%\section{Declaration of interests}
%The authors report no conflict of interest.

\appendix
\section{}\label{appA}
Algorithm~\ref{Algorithm1} presents the Anderson-Moore algorithm with Armijo-type adaptation that can be used to solve an SOF-LQR controller~\citep{Yao2018}.
\begin{center}
\begin{minipage}[t]{0.9\textwidth}
\begin{algorithm}[H] \label{Algorithm1}

\textbf{step 0:} Initialize $F_i=F_0$ to be any $F_0 \in D_s$, where $D_s$ is the set of all stabilizing SOF controllers. Set $0<\xi<1$, $0<\sigma<1/2$, and $\delta>0$. 

\textbf{step 1:} Solve $S(F_i)$ in
\begin{equation*}
    S(F_i)[A+BF_iC]+[A+BF_iC]\trans S(F_i)+C\trans F_i\trans RF_iC+Q = 0.
\end{equation*}

\textbf{step 2:} Set $X_{E}=\mathrm{E}\{X(0)X(0)\trans\}$, solve $H(F_i)$ in
\begin{equation*}
    H(F_i)[A+BF_iC]\trans+[A+BF_iC]H(F_i)+X_E=0. 
\end{equation*}

\textbf{step 3:} Find the smallest integer $\gamma_1 \geq 1$ such that ${F_i+\xi ^{\gamma_1}\mathrm{T}_i} \in D_s$, where
\begin{equation*}
    \mathrm{T}_i = -F_i-R^{-1}[B\trans S(F_i)H(F_i)C\trans ][CH(F_i)C\trans ]^{-1}.
\end{equation*}

\textbf{step 4:} Find the smallest integer $\gamma_M \geq \gamma_1$ such that 
\begin{equation*}J(F_i+\xi ^{\gamma_M}\mathrm{T}_i) \leq J(F_i)+\sigma \xi ^{\gamma_M} \mathrm{trace}(\frac{\partial J}{\partial F_i}\trans \mathrm{T}_i).\end{equation*}

\textbf{step 5:} Find integer $\ell\in\{\gamma_1,\dots,\gamma_M\}$ such that 
\begin{equation*}
J(F_i+\xi ^{\ell}\mathrm{T}_i) =  \min J(F_i+\xi ^{j}\mathrm{T}_i),\text{ where } 
j \in \{ \gamma_1,\cdots, \gamma_M \}.
\end{equation*}

\textbf{step 6:} Set $F_{i+1} = F_i+\xi ^\ell \mathrm{T}_i$, $i = i+1$.

\textbf{step 7:} Check $\Vert \frac{\partial J}{\partial F_i} \Vert _2 \leq \delta$. If true, stop. Otherwise, go to \textbf{step 1}.\\
\caption{\textbf{\textit{Anderson-Moore algorithm with Armijo-type adaptation}}}
\end{algorithm}
\end{minipage}
\end{center}

\bibliographystyle{jfm}
\bibliography{sensor_selection}

\end{document}